# Crowds, Lending, Machine, and Bias

**Runshan Fu**
Heinz College,
Carnegie Mellon University
runshan@cmu.edu

**Yan Huang**
Tepper School of Business,
Carnegie Mellon University
yanhuang@cmu.edu

**Param Vir Singh**
Tepper School of Business,
Carnegie Mellon University
psidhu@cmu.edu

## Abstract

Big data and machine learning (ML) algorithms are key drivers of many fintech innovations. While it may be obvious that replacing humans with machines would increase efficiency, it is not clear whether and where machines can make better decisions than humans. We answer this question in the context of crowd lending, where decisions are traditionally made by a crowd of investors. Using data from Prosper.com, we show that a reasonably sophisticated ML algorithm predicts listing default probability more accurately than crowd investors. The dominance of the machine over the crowd is more pronounced for highly risky listings. We then use the machine to make investment decisions, and find that the machine benefits not only the lenders but also the borrowers. When machine prediction is used to select loans, it leads to a higher rate of return for investors and more funding opportunities for borrowers with few alternative funding options.

We also find suggestive evidence that the machine is biased in gender and race even when it does not use gender and race information as input. We propose a general and effective "debiasing" method that can be applied to any prediction focused ML applications, and demonstrate its use in our context. We show that the debiased ML algorithm, which suffers from lower prediction accuracy, still leads to better investment decisions compared with the crowd. These results indicate that ML can help crowd lending platforms better fulfill the promise of providing access to financial resources to otherwise underserved individuals and ensure fairness in the allocation of these resources.

**Keywords:** fintech, peer-to-peer lending, crowdfunding, machine learning, algorithmic bias

## 1. Introduction

The financial industry is being transformed by new technological innovations. From operational changes such as branchless banking to disruptive new business models such as peer-to-peer (P2P) lending, information technology has been shaping and advancing the industry in various ways (Gomber et al. 2018). One of the driving forces of many fintech innovations and what differentiates the current wave of technology adoption from the previous ones is the use of *big data* and *machine learning* (ML) algorithms.

As we increasingly use machines to facilitate or even replace humans in high-stakes financial decisions such as access to credit, two questions are of particular importance in this space but have not yet been answered. (1) Can machines outperform humans and make better decisions? If so, where and why can machines outperform humans, and what are the economic and welfare implications? (2) Many human decisions are known to be biased (against certain demographic groups). Do machines automate or even amplify such biases, or can they remove the structural inequalities that are found in society, especially in the allocation of financial resources? In this paper, we answer these questions by comparing machine decisions and human decisions in the context of P2P lending.

A key feature of P2P lending is that the investment decisions are collectively made by a crowd of investors, in which case the individual prediction errors might be cancelled out and the wisdom of the crowd can be leveraged. While ML models can be powerful, a market is also hard to beat. The general efficient market hypothesis (Fama 1970) states that prices of assets fully reflect all publicly available information. Built on similar ideas, the prediction market literature (Wolfers and Zitzewitz 2004) suggests that when a crowd of people put their money at stake for an uncertain future outcome, the market could produce an accurate prediction of the outcome (Arrow et al. 2008). P2P lending is a good context to study the performance of ML algorithms compared to the wisdom of a crowd. First, the investment decisions in P2P lending primarily hinge on default risk predictions of loan requests (listings), which makes them typical prediction policy problems, and this is where machines have the potential to assist human decision-making and improve upon human decisions. Second, P2P lending naturally leverages the wisdom of crowds and serves as a prediction market because of the stakes that are involved, even though the implementation may not be perfect due to ineffective system designs such as the herding effect and the one-sided auction rule.

In this paper, we show that despite having access to less information about borrowers, a reasonably sophisticated ML model[1] outperforms crowd investors in predicting listing default risk and leads to better investment decisions that benefit both investors and borrowers simultaneously. Specifically, we build an XGBoost model to predict listing default risk using data from Prosper.com, and then use the machine predicted default risk to create portfolio of listings to invest in. We show that machine created portfolios of loans outperform crowd created portfolios in terms of net present value (NPV) and internal rate of return (IRR).

Comparing machine versus crowd presents unique challenges. First, the machine can be trained only on funded listings (loans) for which we know the outcome (i.e. default/no default). This is known as *selective labels* problem, i.e., "the process generates only a partial labeling of the instances, and the decision-maker's choices determine which instances even have labels at all" (Lakkaraju, et al., 2017). Once the ML algorithm is trained, it may tell us on a test set that some of listings should not be funded based on predicted risk of default. However, the crowd funded all these listings. As a result, by default, the machine should outperform the crowd on prediction accuracy. This comparison is inherently unfair. Second, the risk preferences of the crowd as well as the crowd predicted default risk for the loans is unknown.

We overcome the *selective labels problem*, unobservability of crowd risk preferences and the crowd predicted default risk (Kleinberg et al. 2018). We compare machine versus crowd on the performance of their investment portfolios which are composed of a subset of listings for which outcome labels are available. The loans are selected into the machine and crowd portfolios based on machine predicted default risk and crowd predicted default risk, respectively. We perform two comparisons. In the first comparison, we create portfolios with least risky loans as predicted by machine and crowd. Since crowd predicted default risk is unobserved to us, we use ordinal information provided by the interest rate of the loans. Of any two listings, the crowd predicted default risk is assumed to be higher for the one which is funded at a higher interest rate. We show

---

[1] In the paper, by "sophisticated", we mean "sophisticated in the current literature" or "sophisticated in current times"

that the machine portfolio outperforms the crowd portfolio for any loan amount on NPV and IRR. While this comparison is intuitive, it implicitly assumes risk-neutral crowd, which may not be correct.

The second comparison accounts for the fact that the risk preferences of the crowd are unobserved to us. In this comparison, we divide the crowds into quintiles in such a way that the listings are randomly distributed across the quintiles and that the quintiles differ in their risk (measured by return variance of loans). In this case, each crowd quintile represents an investment portfolio. From the most-risky crowd quintile, we can remove the riskiest loans using the machine predicted default risk until the overall risk is similar to that of any of the other less risky crowd quintiles. In this way, we can fix the risk and compare two portfolios on returns – a pure crowd portfolio and a machine improved crowd portfolio. No matter whether the crowd is risk-averse or risk-seeking, given the same portfolio risk, the crowd will always prefer the portfolio that gives a higher return. We show that for any amount of risk, the machine improved crowd portfolio significantly outperforms the pure crowd portfolio in terms of NPV and IRR.

We further find that the improvement in prediction accuracy (i.e., alignment between the predicted default risks and the true risks) by the algorithm is higher for listings that have high default risk. This can be explained by the design of the P2P platform's auction system. The prevailing interest rate starts from the maximum interest rate that the borrower is willing to accept and can be bid down by interested investors. Thus, a higher than fair interest rate being offered by an investor due to her overestimation of the default risk can be corrected by the lower interest rate that is offered by investors coming later. In contrast, a lower than fair interest rate offer resulting from an underestimation of the default risk cannot be corrected. In cases where the listing default rate is high, the latter scenarios occur more frequently, and, as a result, crowd decisions can be way off.

We also find that borrowers with fewer alternative funding options due to low credit scores, no home ownership or lack of long credit history are more likely to be funded by the machine than by the crowd. Therefore, machine prediction can help P2P lending platforms better deliver their promises to investors and borrowers – including steady returns on investment and more funding opportunities – simultaneously.

Arising with the popularity of ML applications is the concern about (unintended) bias that can be hidden in algorithms, which is particularly relevant for lending markets. If ML algorithms produce biased results, then such social biases would be perpetuated in credit lending decisions, thereby limiting the financial resources that would be available to disadvantaged groups. Even when an ML model does not use a sensitive attribute (e.g., race) as an input, its prediction outcome (e.g., default risk) can still (and often) be discriminatory with respect to a sensitive attribute (e.g., race) if the input features are correlated with the sensitive attribute. In our context, we find suggestive evidence for similar biases in our machine prediction under several commonly used fairness notions.

We propose a general and effective method to remove the bias in ML predictions. This debiasing method is applicable to any prediction focused ML application and causes only minimal accuracy losses in our empirical setting. After being effectively "debiased", the machine still outperforms the crowd in default predictions and investment decisions, which further illustrates the welfare gains that ML algorithms could bring in the context of P2P lending.

The rest of the paper is organized as follows: Section 2 reviews the related literatures, Section 3 introduces our data and analysis context, Section 4 compares machine predictions and crowd decisions, Section 5 examines the potential bias in machine and crowd predictions and presents our debias method, and Section 6 discusses the managerial implications of our results and concludes the paper.

## 2. Literature Review

Our paper is related to multiple streams of literature in fintech and computational social science. First, it contributes to the literature on P2P lending, which is an important fintech innovation. Iyer et al. (2016) documented the effectiveness of market screening on P2P lending platforms. They showed that peer lenders' predictions of a borrower's likelihood of defaulting were 45% more accurate than the predictions based on the borrower's exact credit score, which was not observed by the lenders.[2] Our analysis shows consistent results, and, more importantly, demonstrates that an ML algorithm, even with access to less listing information, can outperform the crowd of lenders in predicting listing default risk and further improve returns on investment. A number of papers have examined user behaviors on P2P lending and crowdfunding platforms. On the investors' side, previous research has identified factors affecting their funding decisions, such as peer lenders' decisions (Liu et al. 2015, Zhang and Liu 2012), borrowers' identity claims (Herzenstein et al. 2011), appearance (Duarte et al. 2012), friendship networks (Lin et al. 2013, Liu et al. 2015), proximity to the lenders' location (Lin and Viswanathan 2016), and tax incentives (Chen et al. 2018). While some of these factors help investors more accurately estimate default risk and make better investment decisions, others do not and sometimes prevent investors from making the optimal decisions. In this paper, we use a predictive algorithm as a calibration and diagnostic tool and show that investors' *collective* predictions are fairly accurate for low risk listings; however, their predictions are less accurate for high risk listings, and the inaccuracy can be attributed to over- or under-weighing risk indicators and an imperfect auction system design. On the borrowers' side, Butler et al. (2017) study how local capital market conditions affect the interest rates that borrowers are willing to accept on crowdlending platforms; also, Kawai et al. (2014) showed that borrowers can use a reserve interest rate to signal their creditworthiness, and this signal can effectively restore the damage that is caused by adverse selection (i.e., "market of lemon" – only low-quality listings are listed).

---
[2] On Prosper.com, lenders only observe the interval in which borrows' credit score falls, but not the exact credit score.

Our paper studies the impacts of the using ML algorithm to select loans on borrowers' welfare too, and finds that, compared with peer investors' decisions, investment decisions based on the ML algorithm give more opportunities to borrowers with fewer alternative funding sources. Gomber et al. (2018) identifies such borrowers as less "bankable" individuals and the primary beneficiaries of P2P lending. Taken together, we show that a reasonably sophisticated ML algorithm can help P2P lending platforms better deliver benefits to both investors and borrowers simultaneously.

There have been a number of studies in the fintech literature on the use of data mining and ML in fintech applications, such as stock price/return predictions (Gu et al. 2018), risk management (Butaru et al. 2016, Khandani et al. 2010, Netzer et al. 2019), and fraud prevention (Fawcett and Foster Provost 1997). However, most of these works focus on identifying useful features and/or developing better models to improve the prediction accuracy. Our work differs from this stream of literature in that it goes beyond *predictions* to study *decision-making* based on machine predictions and its social and economic implications.

The problem that we study in this paper is a *prediction policy problem* (Kleinberg et al. 2015). In the growing literature on prediction policy problems, the use of ML models and their potential interplay with humans have been examined in the context of hiring (Chalfin et al. 2016), surgery allocation (Kleinberg et al. 2015) and bail decisions (Kleinberg et al. 2017). These papers generally find that the use of an ML algorithm can improve individuals' or experts' decisions. This paper studies the application of ML models in lending decisions, which is a high-potential application area for ML models in finance; furthermore, uniquely, to the best of our knowledge, it is the first that compares machine predictions with crowd decisions, which has been shown in the prediction market literature (e.g., Qiu and Kumar 2017) to be able to produce accurate predictions of uncertain outcomes when individuals are knowledgeable and motivated to be accurate.

Methodologically, this paper relates to the stream of literature on fairness in algorithms. Recently, the fairness issues of ML algorithms have attracted attention from researchers in the ML community, and many fairness notions have been proposed and discussed (Chouldechova and Roth 2018, Hardt et al. 2016). We adopted several of these notions to measure bias in our research context to quantify and compare the crowd bias and the machine bias. We find that both the crowd and the fairness-unaware machine are biased no matter what fairness notion is used. ML researchers have also proposed several methods for mitigating or eliminating bias in ML algorithms (Berk et al. 2017, Calders et al. 2009, Hardt et al. 2016, Kamiran et al. 2010, Pedreshi et al. 2008), and demonstrate the effectiveness and efficiency of removing or reducing biases for specific applications. However, the existing debiasing methods have drawbacks such as (1) being limited to certain ML algorithms (e.g., tree-based model); (2) being limited to certain prediction tasks (e.g., binary classification); and (3) requiring changing outcome labels, which is justified in some applications where the label itself is subjective, but not in cases where the label is objective (as in our case). We propose a debiasing method and explain in detail its application to various types of variables (e.g., continuous, ordinal, nominal and mixed variables). We further demonstrate the use of the proposed debiasing method in our P2P lending context and empirically show that the method is effective in removing bias and that the accuracy loss resulting from the debiasing method is small. The proposed method addresses several major drawbacks of these existing methods that are mentioned above and is general to almost any machine prediction application contexts.

## 3. Data and Context

Our primary research context is the Prosper Marketplace (Prosper.com), which is the first peer-to-peer (P2P) lending marketplace in the United States. Our main dataset consists of loan requests (listings) that have been posted on Prosper and repayment performance of all the successfully funded loans between March 2007 and October 2008. We choose data in this period because there was minimal loan selection on Prosper's side,[3] and the platform operated an auction-based pricing model that allowed investors to bid on interest rates.[4]

Under the auction-based pricing model, a borrower submits a loan request specifying the loan amount and the maximum interest rate she would accept, along with several financial characteristics. All the information is available to lenders while they browse the listing. If a lender is interested in the listing, she can bid on it by specifying the amount she wants to contribute (minimum $50) and the lowest interest rate she would accept. The prevailing interest rate for a listing starts from the maximum borrower rate and decreases when the investors' pledged amount exceeds the borrower's requested amount.

In the dataset, there are 247,443 listings in total. The listing data file contains most of the information that is available to the investors when they browsed a listing, including the borrower's location (city level), employment status, occupation, debt-to-income ratio, homeownership, credit score level, inquiries in the last 6 months, and several other credit features. Among all the listings, 20,668[5] (approximately 8.3%) were successfully funded. For the funded loans, we observe the final interest rate and their repayment performance, which includes the amount of paid principal, the amount of paid interest, and the loan status (charge off, defaulted, or completed).

---

[3] In October 2008, Prosper was shut down and started a registration process with the U.S. Securities and Exchange Commission. When it re-launched in July 2009, Prosper made several significant changes, including a minimum credit score requirement for borrowers. It added more borrower eligibility criteria in the subsequent years.
[4] In December, 2010, Proper switched to a posted price model in which the interest rate is solely determined by Prosper based on its evaluation of each prospective borrower's credit risk.
[5] Those that were initially funded but later cancelled are excluded.

Prosper masks the listing identifier in the loan performance data file; therefore, we cannot directly merge the two data files. Instead, we match the loans in the two files based on their requested loan amounts, final interest rates and loan origination dates. This allows us to uniquely identify 19,529 funded loans and combine their listing characteristics with the repayment performance.[6] These **funded loans with matched performance labels** form our main dataset for training the ML algorithm and performing the crowd vs machine comparison, because we need to know the outcome (default/No default) for model training and comparison purposes. Table 1 displays the summary statistics of the unfunded listings, the funded loans, the completed loans and the defaulted loans.[7]

**Table 1. Summary Statistics of the dataset**

|  | Unfunded | Funded | Completed | Defaulted |
|---|---|---|---|---|
| Sample size | 226775 | 20668 | 13286 | 6243 |
| **Listing characteristics** | | | | |
| Listing amount | 8001.13 | 6540.63 | 6271.63 | 7371.89 |
| Listing term | 36 | 36 | 36 | 36 |
| Monthly payment | 296.17 | 234.00 | 220.38 | 271.95 |
| Interest Rate Paid (%) | - | 18.00 | 16.35 | 20.74 |
| Interest Rate Max (%) | 20.67 | 21.65 | 20.12 | 24.27 |
| **Borrower Characteristics** | | | | |
| Stated monthly income | 4140.59 | 4637.69 | 4715.33 | 4558.22 |
| Income verifiable | 0.89 | 0.94 | 0.95 | 0.93 |
| Debt to Income ratio | 0.51 | 0.32 | 0.29 | 0.39 |
| Months employed | 65.30 | 68.95 | 67.49 | 72.13 |
| Employment status | | | | |
| *Full-time* | 0.78 | 0.86 | 0.86 | 0.84 |
| *Self-employed* | 0.10 | 0.08 | 0.07 | 0.10 |
| *Part-time* | 0.04 | 0.04 | 0.04 | 0.03 |
| *Retired* | 0.03 | 0.02 | 0.02 | 0.02 |
| *Not employed* | 0.02 | 0.01 | 0.01 | 0.01 |
| *Employed* | 0.03 | 0.00 | 0.00 | 0.00 |
| *Other* | 0.002 | 0.00 | 0.00 | 0.00 |
| Has prior Prosper loans | 0.05 | 0.11 | 0.11 | 0.10 |
| Is homeowner | 0.35 | 0.47 | 0.46 | 0.51 |
| Is Prosper lender | 0.10 | 0.25 | 0.29 | 0.17 |
| Number of Public Records (last 10 years) | 0.60 | 0.36 | 0.30 | 0.46 |
| Number of Public Records (last 12 months) | 0.06 | 0.03 | 0.03 | 0.04 |
| Credit Characteristics | | | | |
| Credit Grade | | | | |
| *AA* | 0.40 | 0.08 | 0.06 | 0.13 |
| *A* | 0.18 | 0.08 | 0.07 | 0.11 |

---

[6] When two or more loans have identical loan amounts, final interest rates and loan origination dates but different repayment performance, we cannot match the loans with their repayment performance.
[7] Table A.1 in the Online Appendix A1 displays the summary statistics of the full sample (all listings), the funded set and the funded loans with matched labels. It shows that the funded set and the funded set with matched labels are not statistically different in their characteristics.

|  |  |  |  |  |  |
|---|---|---:|---:|---:|---:|
|  | *B* | 0.17 | 0.18 | 0.17 | 0.20 |
|  | *C* | 0.12 | 0.21 | 0.21 | 0.22 |
|  | *D* | 0.06 | 0.17 | 0.18 | 0.16 |
|  | *E* | 0.04 | 0.13 | 0.15 | 0.10 |
|  | *HR* | 0.03 | 0.13 | 0.16 | 0.07 |
| ScoreX |  |  |  |  |  |
|  | *< 600* | 0.24 | 0.17 | 0.13 | 0.24 |
|  | *600-619* | 0.05 | 0.09 | 0.08 | 0.11 |
|  | *620-639* | 0.05 | 0.09 | 0.09 | 0.10 |
|  | *640-649* | 0.02 | 0.07 | 0.07 | 0.08 |
|  | *650-664* | 0.03 | 0.08 | 0.08 | 0.09 |
|  | *665-689* | 0.03 | 0.11 | 0.12 | 0.11 |
|  | *690-701* | 0.01 | 0.05 | 0.05 | 0.05 |
|  | *702-723* | 0.02 | 0.08 | 0.09 | 0.07 |
|  | *724-747* | 0.02 | 0.08 | 0.10 | 0.06 |
|  | *748-777* | 0.01 | 0.08 | 0.09 | 0.06 |
|  | *778+* | 0.01 | 0.08 | 0.11 | 0.03 |
|  | *Missing* | 0.51 | 0.00 | 0.00 | 0.00 |
| Current credit lines |  | 8.16 | 9.58 | 9.57 | 9.71 |
| Open credit lines |  | 7.21 | 8.22 | 8.20 | 8.35 |
| Bank utilization |  | 0.64 | 0.55 | 0.52 | 0.60 |
| Total open revolving accounts |  | 5.88 | 6.36 | 6.44 | 6.26 |
| Installment balance |  | 29237.55 | 26158.08 | 24978.05 | 28545.96 |
| Real estate balance |  | 92917.28 | 124228.62 | 112241.83 | 155948.95 |
| Revolving balance |  | 17675.10 | 19061.02 | 18338.84 | 21077.87 |
| Total inquiries |  | 11.96 | 9.04 | 7.61 | 12.09 |
| Inquiries in last 6 months |  | 4.07 | 2.60 | 2.06 | 3.72 |
| Total trade items |  | 26.12 | 24.60 | 23.94 | 26.13 |
| Satisfactory accounts |  | 18.98 | 20.26 | 20.12 | 20.83 |
| Now delinquent derog |  | 2.70 | 0.96 | 0.70 | 1.47 |
| Was delinquent derog |  | 4.43 | 3.38 | 3.11 | 3.83 |
| Delinquencies over 30 days |  | 9.96 | 6.67 | 6.01 | 7.87 |
| Delinquencies over 60 days |  | 4.67 | 2.75 | 2.42 | 3.34 |
| Delinquencies over 90 days |  | 9.44 | 4.93 | 4.27 | 6.11 |
| Amount delinquent |  | 3671.10 | 1100.30 | 848.24 | 1576.84 |
| Length of credit history |  | 4800.73 | 4907.40 | 4878.97 | 5006.79 |
| **Outcomes** |  |  |  |  |  |
| Default |  |  | 0.32 | 0.00 | 1.00 |
| Principal Paid |  |  | 4898.07 | 6192.82 | 2263.72 |
| Interest Paid |  |  | 1341.93 | 1359.48 | 1336.61 |

The original dataset does not contain any sensitive demographic features such as age, gender or race. To measure the bias in machine prediction, we use borrowers' occupation and location information to create approximate gender groups and race groups. Specifically, we cross reference the borrowers' occupations with the 2008 Current Population Survey (CPS) data and assign listings submitted by borrowers who were in high female concentrated occupations (female percentage greater than 75%) into the 'female' group, and those by borrowers in low female concentrated occupations (female percentage less than 25%) into the 'male' group. Similarly, we map each borrower city to a unique Federal Information Processing Standard Publication (FIPS) code and pull the demographic information for all the FIPS areas from the 2010 Decennial Census data. The listings submitted by borrowers who lived in areas with relatively high saturations (the top 25 percentile) of black people are assigned to the 'black' group, and those that were requested by borrowers who lived in relatively low black saturated areas (the bottom 25 percentile) are assigned to the 'nonblack' group. Table A.2 and Table A.3 in the Online Appendix A1 present the summary statistics for the gender groups (female vs. male) and race groups (black vs. nonblack), respectively. Our demographic group assignments are approximate, yet the differences between the groups are evident. In our sample, on average, the male borrowers have better credit profiles than the female borrowers (higher incomes, lower DTI ratios, fewer public records, higher credit scores, etc.), while the black borrowers have better credit profiles than the nonblack borrowers.

## 4. Machine Prediction vs. Crowd Decision

In this section, we will first train an ML algorithm and then evaluate its performance. The training process is similar to the standard supervised ML practice. Given a dataset of several characteristics (listing and borrower characteristics) and an outcome label (default or not), we train a model that learns the mapping from the features to the label. As mentioned earlier, our focus is on the evaluation part. Instead of simply measuring the prediction accuracy of the trained algorithm using out of sample data (test set), we are more interested in evaluating whether machine predictions can improve investment decisions that result in welfare gains for investors and borrowers. We randomly split the main dataset defined in the previous section into a training set that contains 60% of the data (11,717 loans), which will be used to train the machine, and a test set that contains 40% of the data (7,812 loans) which will be used to perform the crowd vs. machine comparison.[8]

We would like to reiterate that the ML algorithm is trained and comparison is performed on the funded set in which we have the outcome labels, not the full sample of listings. For the comparison, we follow Lakkaraju et al. (2017) and Kleinberg et al. (2018), which develop methods for machine and human fair comparison in the presence of selective labels.

---

[8] In an additional analysis (presented in the Online Appendix A5), we divide our sample into 4 time periods with equal length and build a machine learning model for each of the periods to examine the potential shift in lenders' and borrowers' behavior. In another additional analysis (presented in the Online Appendix A7), we consider a machine learning model with a rolling window.

### 4.1 ML Model to Predict Listing Default Risk

We train a ML model that uses available features of the listing to predict listing default risk based on the training set described above.[9] Our outcome of interest (Y) is loan default, which is a binary variable that equals 1 if the loan status is "default" or "charge off", and 0 if the status is "complete". The input features (X) include the loan amount, borrower characteristics and their credit information. We exclude the final interest rate and monthly payment from the feature list because these are the crowd investors' decisions and the outcomes of the funded loans and should not be made available to the machine as input features (this information is not available when the machine makes the decision on whether to invest in a listing). We further exclude borrowers' locations and occupations because they are used to proxy for race and gender, and the current law prohibits explicitly considering race and gender information when making investment decisions.[10] As we will show later in Section 4.3, the machine we build, with an objective of predicting default risk and excludes some variables that may be predictive of the default risk (e.g., the final interest rate) for reasons mentioned above, can still outperform the crowd in terms of overall returns.

We use an XGBoost model (Chen and Guestrin, 2016) to fit a prediction function $\phi(X)$ that outputs the predicted probability of default $P(Y = 1|X)$. As previously mentioned, the machine is trained on the funded set; therefore, the prediction generated by the machine is $\phi(X|R^* = 1)$, instead of $\phi(X)$. XGBoost is a scalable tree based boosting system that achieves current state-of-the-art results in many ML challenges, including many finance applications such as stock market event forecasts (Chatzis et al. 2018) and credit risk assessment (Nguyen, 2019). An XGBoost model is a tree ensemble model that consists of multiple regression trees (also known as CARTs). Unlike normal decision trees that output class labels or use the portion of positive classes as the class probability, a regression tree performs a regression in each leaf node. In our case, the logistic regression is used and each tree outputs the probability of default. The final prediction of an XGBoost model is the sum of the predictions from each regression tree. To learn a model, we minimize an objective function that consists of the training loss and regularization term. Intuitively, we aim to balance the prediction accuracy and the model's simplicity since we minimize the sum of the training errors and the model's complexity. Tree based models are vulnerable to overfitting problems. In practice, we use several techniques to alleviate it. First, when fitting regression trees, we can specify a maximum depth so that a tree stops growing once it reaches the depth. Second, we can subsample instances and/or features and therefore create slightly different datasets for each tree. Third, after we learn a tree, we usually shrink it when adding it to the model.

---

[9] Recall that all listings in the training set are funded loans, because default status is only available for funded loans.
[10] The inclusion of the four features would increase the prediction accuracy.

There are several hyperparameters in an XGBoost model: the number of trees, the regularization terms, the learning rate, the maximum tree depth, the instance subsample percentage and the feature subsample percentage. We tune these parameters using five-fold cross-validation on the training set. Because our hyperparameter space is relatively large and fitting the XGBoost model is computationally expensive, we choose the Bayesian Optimization instead of the more commonly used grid search or random search to search for the optimal hyperparameters. The final tree depth we choose is eight. Details of the XGBoost are provided in Online Appendix A2.

## 4.2 Crowd Prediction of Default Risk

We do not directly observe the crowd predicted default risk in the data. Instead, we use risk premium to proxy for it. The rationale is as follows. Under an auction-based market mechanism, crowd investors on Prosper collectively made two decisions for each listing: the funding decision $R \in \{0, 1\}$, and the interest rate decision $r \in [0, 0.35]$. Investors observed all the input features ($X$) of our XGBoost model, and they also observed several other features that are unobservable to us (such as text descriptions or borrowers' profile images) or not used by the model (such as borrowers' locations or borrowers' occupations). We denote these features as $Z$. We assume that investors made decisions based on their assessment of the listing default risk $m(X, Z)$. They intend to fund listings and, among these listings, they would like to charge higher interest rates for listings with higher default risks as follows:

$$R = 1 \; iff \; m(X, Z) < \alpha \tag{1}$$

$$r(X, Z | R = 1) = g(m(X, Z | R = 1)) + c(X, Z | R = 1) \tag{2}$$

where $\alpha$ is the funding threshold, $g(\cdot)$ is a strictly increasing function, and $c(X, Z|R = 1)$ is the part of the interest rate that is contributed by other factors (other than predicted default risk). Notice that the final funding outcome ($R^* \in \{0, 1\}$) requires an agreement between a borrower and investors. A listing would be successfully funded only when the investors intended to fund it and the interest rate was acceptable to the borrower as follows:

$$R^* = 1 \; iff \; R = 1, r(R = 1) \leq b \tag{3}$$

where $b$ is the borrower's specified maximum interest rate.

It is tempting to interpret a binary funding outcome $R^*$ as the crowd prediction of the default risk and compare its accuracy with a funding decision that is derived from $m(X, Z)$. However, there are two problems with this approach. First, $R^* = 0$ does not necessarily mean that crowd investors predicted that the loan would end in default ($R = 0$). It could be the result of an interest rate mismatch, i.e., $r(R = 1) > b$. Thus, the $m(X, Z)$ of a funded listing could be higher than that of an unfunded listing. Ignoring this fact would underestimate the accuracy of crowd predictions and lead us to a biased conclusion that favors machine predictions. Second, we only observe the outcome label ($Y$) for listings that were funded ($R^* = 1$), but do not observe it for unfunded listings ($R^* = 0$). Consequently, we cannot compute the accuracy of crowd decisions, and the accuracy of machine predictions can only be computed for a selected sample.

To get around these problems, we focus on the funded set in which we have the outcome labels. Among all the funded loans, those that were given a higher interest rate arguably had been predicted as riskier, and we utilize this information to evaluate crowd predictions. Specifically, we define the risk premium as follows:

$$p(X, Z | R^* = 1) = r(X, Z | R^* = 1) - rf_t \tag{4}$$

where $rf_t$ is the risk-free rate (3-Year Treasury bill rate) at the time of the loan request. We use $p(X, Z| R^* = 1)$ as a proxy for $m(X, Z|R^* = 1)$.[11] The rationale is explained below.

Default risk is one of the most important components of interest rates for commercial loans (Cox et al. 1985, Merton 1974). Obviously, investors expect to be properly compensated for bearing the risk that they may not get their money back in the event of a loan default. In other words, investors ask for a higher return as compensation for a higher (perceived) probability of investment failure.

---

[11] In an alternative specification, we assume that the investors used a fixed risk-free rate throughout the time period and perform the same analysis. The results are presented in the Online Appendix A4.

The risk-free rate is the baseline for any interest rate. It is the theoretical rate of return for an investment that has no risk, and treasury bills are commonly used as the reference risk-free rate because their default risk is negligible. Everything else equal, the interest rate is higher when the risk-free rate is higher.

Default risk and risk-free rate fluctuations are two major contributors to the difference in interest rates among the funded loans.[12] Other components of the interest rate were constant for all the listings in our data. For instance, all the listings were 3-year fixed rate unsecured personal loan requests; therefore, they had the same maturity premium. Furthermore, no secondary market for Prosper investment bills was available; thus, the same liquidity premium (Liu et al. 2006) applied to all. Other factors such as accounting transparency and tax factors were not different for any of the listings. Based on the above reasoning, the interest rate for the funded loans on Prosper can be written as follows:

$$r(X,Z|R^* = 1) = g(m(X,Z|R^* = 1)) + rf_t + c \qquad (5)$$

where $c$ is a constant. Thus, the corresponding risk premium can be written as follows:

$$p(X,Z|R^* = 1) = r(X,Z|R^* = 1) - rf_t = g(m(X,Z|R^* = 1)) + c \qquad (6)$$

It is worth noting that the value of $g(m(X,Z|R^* = 1))$, the varying part of the risk premium, is affected by two factors: the crowd estimated default risk, $m(X,Z|R^* = 1)$; and the risk preference. Individual investors may have different risk preferences and different projects may have different pools of investors. Therefore, the estimated risk and the risk premium is not perfectly correlated, i.e., $g(\cdot)$ is not necessarily linear. Our basic assumption is that an investor always assigns a higher interest rate to a riskier listing because she needs to be compensated for bearing the additional risk. This implies that $g(\cdot)$ is an arbitrary increasing function. Therefore, $p(X,Z|R^* = 1)$ and $m(X,Z|R^* = 1)$ have the same rank order. Later, when comparing machine decisions and crowd decisions, we rely on only the rank orders of $m(X,Z|R^* = 1)$ instead of the magnitudes of the predicted risks to discriminate the listings to invest in for our first comparison. In this sense, $p(X, Z| R^* = 1)$ can be used as a proxy of $m(X, Z|R^* = 1)$.

---

[12] In the post-price period, when Prosper sets the interest rate, the loss rate explains the majority of the variation in interest rates across different loans, which supports this claim.

### 4.3. Machine vs. Crowd Comparison

### 4.3.1 Prediction Accuracy

We first use a common metric, the Area Under the Receiver Operating Characteristic curve (AUC-ROC, to measure the prediction accuracy. The AUC-ROC is an evaluation metric that is good for measuring predicted score performance when $Y$ is unbalanced, as in our case where there are more completed loans ($Y = 0$) than default loans ($Y = 1$). The AUC-ROC for the machine and the crowd are 0.741 and 0.678, respectively. (The details of how to generate the AUC curves for the machine and the crowd are provided in Online Appendix A3.) This suggests that, overall, our ML model has more predictive power than the crowd.

To understand the performance of the predictions at a more detailed level, we also check the distributions of the predictions. We bin the loans in the test set into 100 quantile groups based on the machine predicted risk $\phi(X|R^* = 1)$. Figure 1 plots the actual default rate and the average risk premium against the average machine predicted risk in each group in the two panels, respectively.

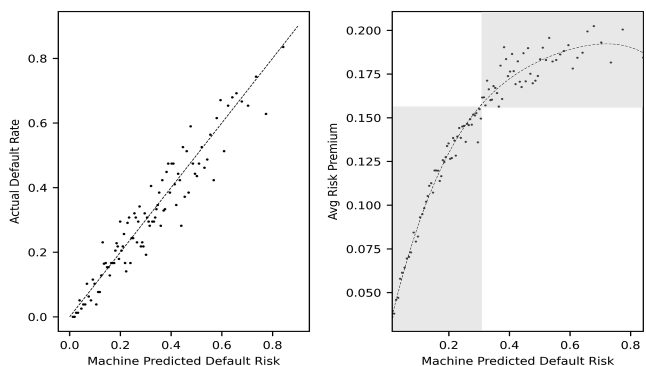

**Figure 1 Distribution of Machine Predicted Risk**

The left panel shows that our ML model is well calibrated. The data points are centered around the line $y = x$ over the entire risk distribution, suggesting that the machine predicted risk is aligned with the true risk. In the right panel, we can see that the machine prediction and the crowd prediction are highly correlated for low risk loans ($\phi < 0.3$). In the low risk range, the points in the right panel are clustered more tightly around the fitted curve, which suggests that the model and the crowd have extracted similar predictable risk from the observables. (The fitted curve does not necessarily have a 45-degree angle because the actual risk may include parts that are unpredictable or cannot be inferred from the available data.) However, for high risk loans, the ML model and the crowd tend to disagree. For loans with a machine predicted risk ranging from 0.3 ("will probably default") to 0.8 ("very likely to default"), the average risk premium only ranges between 0.15 to 0.20 as if the crowd cannot distinguish the high-risk loans from the medium-risk loans. Note that the left panel suggests that the loans that the machine predicts as risky are indeed risky loans. Thus, this means that the crowd predictions fall short for the relatively high-risk loans.

Overall, these results suggest that the ML model has more predictive power than the crowd. However, the machine does not always outperform the crowd. Specifically, the crowd predictions are consistent with the machine predictions for low risk loans but are inaccurate for high risk loans.

### 4.3.2 Decision Quality -- Welfare Analysis

In the previous section, we show that our ML model predicts default risk more accurately than the crowd, but it is not obvious how much economic benefit this improved accuracy brings. In this section, we compare machine decisions and crowd decisions in terms of investor and borrower welfare. The comparison is done on the test set.

**Investors: Higher Return**

We perform two comparisons to show that the machine can improve upon crowd investment decisions and generate higher investor welfare. In both comparisons, we follow the framework proposed by Lakkaraju et al. (2017) and Kleinberg et al. (2018) to address the selective labels problem previously discussed.

*Comparison (1)*

For this comparison, we define the investors' utility as the NPV of all the loans they invest in:

$$u(D) = \sum_{i \in D}(-a_i + \sum_{t=1}^{T}\frac{w_{it}}{(1+\alpha)^t}) \tag{7}$$

where $D$ is an investment portfolio, and $a_i$ is the amount that is funded for loan $i$ in the portfolio. $T$ is the term of the loans in months (36), $w_{it}$ is the payback amount for loan $i$ in month $t$ (cashflow), and $\alpha$ is the discount rate. We do not observe the exact cashflows for the loans. Following Lin et al. (2019), we first calculate the repayment schedule for each loan based on the loan amount, interest rate, and loan term, which are observable to us. We use the repayment schedule to proxy for the actual cashflows for the paid-in-full loans. For the defaulted loans, we use the repayment schedule and the amount principal repaid to determine when the loan defaulted, and then compute the cashflows until default.[13] We use the average risk-free rate during the period (2.8%/12) as the discount rate.[14]

Another common metric for evaluating investment portfolios is the Internal Rate of Return (IRR). The IRR is a discount rate that makes the NPV of a portfolio equal to 0. I.e., the IRR is the $\alpha$ that satisfies the following:

$$\sum_{i \in D} -a_i + \sum_{t=1}^{T} \frac{\sum_{i \in D} w_{it}}{(1+\alpha)^t} = 0 \tag{8}$$

For an investment portfolio, we first construct the total cashflows by aggregating the cashflows of the individual loans in the portfolio, and then calculate the IRR for the portfolio.[15] A high IRR suggests that the portfolio has a high profitability or high rate of return on investment.

We denote the market portfolio that was collectively selected by the crowd investors as $D_C$. As mentioned earlier, we have the *selective labels* problem that the loan performance of unfunded listings is not observed; thus, we cannot compute the counterfactual return of any $D$ that contains listings that do not belong to $D_C$. However, since the problem is one-sided, we have no problem computing the counterfactual return of $D$ when $D \subseteq D_C$.

In the following analysis, we consider loans in the test set as "listings" and construct portfolios from them based on the machine risk predictions and the crowd risk predictions separately. Given a predicted risk score ($\phi$ or $p$), we can sort those "listings" in increasing default risk. Let $l_i^\phi$ be listing $i$'s ranking according to the machine predicted risk, and $l_i^p$ be listing $i$'s ranking according to the crowd predicted risk (risk premium). Given an investment amount $Q$, we can construct a loan portfolio based on the machine predicted risk (denoted as $D_\phi(Q)$) and another based on risk premium (denoted as $D_p(Q)$) as follows:

$$D_\phi(Q) = \{i | l_i^\phi \leq \alpha_\phi\}, \text{ where } \alpha_\phi \text{ satisfies } \sum_{i \in \{i | l_i^\phi \leq \alpha_\phi\}} q_i \leq Q \text{ and } \sum_{i \in \{i | l_i^\phi \leq \alpha_\phi + 1\}} q_i > Q$$

$$D_p(Q) = \{i | l_i^p \leq \alpha_p\}, \text{ where } \alpha_p \text{ satisfies } \sum_{i \in \{i | l_i^p \leq \alpha_p\}} q_i \leq Q \text{ and } \sum_{i \in \{i | l_i^p \leq \alpha_p + 1\}} q_i > Q \qquad (9)$$

---

[13] With limited information on the payment history, we cannot restore the precise cashflows. This is our best effort. We also computed the plain return on investment for different portfolios and have obtained the same qualitative results.
[14] Please note that 2.8% is the annualized risk-free rate. We divide it by 12 to obtain the monthly free rate and use it as the discount rate.
[15] For easy comparison and readability, we report the annualized IRR throughout the paper.

In the equations above, $q_i$ is the amount of loan $i$, and $\alpha_\phi (\alpha_p)$ is the rank threshold based on the machine (crowd) evaluation of the loan risk. The intuition behind the equations is that $D_\phi(Q)(D_p(Q))$ consists of $\alpha_\phi(\alpha_p)$ least risky loans based on the machine (crowd) prediction, where $\alpha_\phi(\alpha_p)$ is determined by the two conditions that the total amount of $\alpha_\phi(\alpha_p)$ least risky loans is less than the investment budget $Q$, and the total amount of $\alpha_\phi + 1$ ($\alpha_p + 1$) least risky loans exceeds $Q$. When $Q$ equals 0, both $D_\phi(Q)$ and $D_p(Q)$ are empty portfolios. As $Q$ increases, we become less conservative and include listings with higher predicted risks in the portfolios. Finally, when $Q$ equals $31,943,854.78 (the total investment amount observed in the test set, and both $D_\phi(t)$ and $D_p(t)$ equal the market portfolio, $D_C$. Note that here we use only the ordinal information of the predicted risks. As the investment amount $Q$ increases, we are gradually expanding the portfolios by adding the next safest listing until we include all the listings in the market portfolio.

To further illustrate the comparison process, consider a hypothetical example where there are a total of five loans in the test set. Assume that the loan amount, crowd predicted risk (i.e., the calculated risk premium) and machine predicted risk of the five hypothetical loans are as shown in Table 2. We reorder the loans by the machine predicted risk and by the crowd predicted risk, as shown in Table 3.

| \multicolumn{4}{c|}{Table 2. Sample Funded Loans} | \multicolumn{2}{c}{Table 3. Sample Funded Loans Reordered} |

| Loans | Amount | Risk Premium* | Machine Predicted Risk | Crowd | Machine |
|---|---|---|---|---|---|
| 1 | $1000 | 19% | 25% | Loan 3 ($2000) | Loan 2 ($2500) |
| 2 | $2500 | 25% | 15% | Loan 1 ($1000) | Loan 3 ($2000) |
| 3 | $2000 | 15% | 18% | Loan 5 ($2500) | Loan 4 ($1000) |
| 4 | $1000 | 21% | 19% | Loan 4 ($1000) | Loan 5 ($2500) |
| 5 | $2500 | 20% | 20% | Loan 2 ($2500) | Loan 1 ($1000) |

*Note: A proxy for crowd predicted risk.*

We then consider these five loans as five "listings" exposed to investors, and construct investment portfolios based on the crowd predicted risk and machine predicted risk from these listings under a fixed budget. For example, if the total amount $Q$ is $5500, then the crowd's choice is loans 3, 1 and 5, while the machine would choose loans 2, 3, and 4. We then evaluate which portfolio, the one the crowd picks or the one the machine picks, gives higher returns based on their realized repayment outcomes.

The left panel of Figure 2 shows the investments and returns (NPV) for the different portfolios that are created using the holdout test set. The leftmost point (the origin) represents the empty portfolio constructed given the investment amount $Q$ of zero – zero investment and zero return. As the total investment amount increases, we include riskier loans into the investment portfolio. The NPV of the portfolios could go either way. That is, it could increase if we include "good" loans that will be paid in full and bring returns and decrease if we include "bad" loans that will end in default and bring losses. When we include all the loans in the test set, we reach the right most point, which corresponds to the portfolio that includes all the loans in the test set – 7812 loans, with a total investment of $51,342,051.92 and an NPV of $-3,567,424.43.

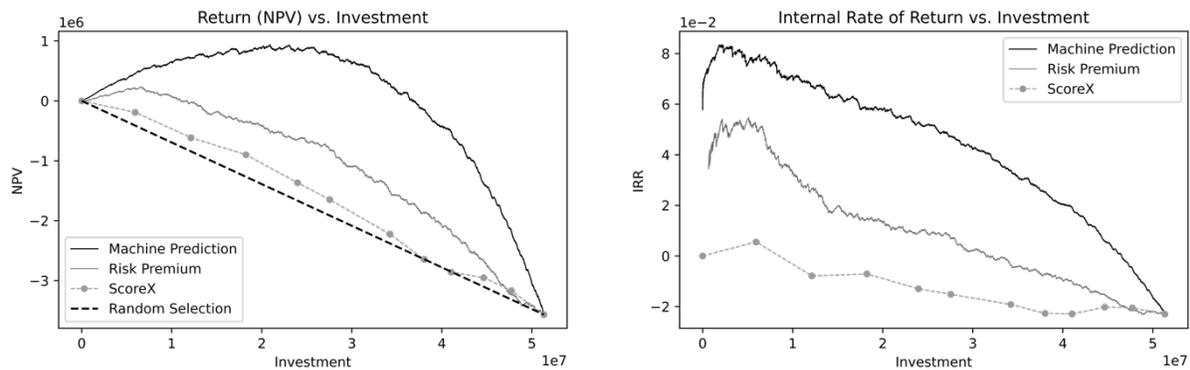

**Figure 1 Return-Investment Plot for the Different Portfolios (Left Panel: NPV; Right Panel: IRR)**

For reference, we also create subset portfolios based on borrower credit score bins (ScoreX).[16] We first include loans whose borrowers are in the highest credit score bin, and then add loans whose borrowers are in the second highest credit score bin. We repeat the process until the portfolio includes all the loans that have been funded. If we draw a vertical line at any point on the x-axis of the left panel of Figure 2, we fix the investment amount, and the intersections of the vertical line with the three curves represent the returns at that investment amount if we pick loans based on machine predicted risk, risk premium (a proxy for crowd predicted risk), and ScoreX, respectively. Finally, we include a curve that shows how the theoretical investment and return change if we randomly add loans.

---

[16] ScoreX is one type of credit scores. Prosper provided the borrower's ScoreX range for each listing (see the summary statistics). There are 11 different bins, and we encoded them as follows: {< 600: 1, 600-619: 2, … 748-777:10, 778+:11}.

The right panel of Figure 2 plots the investments and IRR for the portfolios.[17] This plot is generated with the same process -- we start from an empty portfolio with zero investment,[18] and gradually include more loans until we include all the loans in the test set.[19]

From the plots, we can see that the portfolios based on the ScoreX bins ($D_s$) perform slightly better than random selection, while most portfolios based on the risk premium ($D_p$) significantly outperform those in $D_s$. $D_p$ performs worse than $D_s$ in a few cases when adding presumably high-risk loans (towards the right most point). This further confirms our previous finding that crowd investors were not good at assessing high-risk loans.[20] The return-investment curve of portfolios based on machine predictions ($D_\phi$) strictly dominates the curve based on $D_p$. That is, for any fixed amount of investment, $D_\phi$ provides much higher returns than $D_p$.

These results demonstrate that while crowd predictions are fairly accurate and can lead to substantially higher returns than the industry standard credit score, machine predictions can further improve investment decisions and significantly increase investor welfare. Though the difference in the ROC-AUC of machine predictions and crowd predictions is modest (0.741 vs. 0.678), the impact on financial returns is surprisingly large. To characterize the improvement in returns that machine prediction can provide, we calculate ($NPV^{machine} - NPV^{crowd}$)/$Investment$ for all possible investment amounts from $0 to $51,342,051.92, and find that the largest value is 0.1841, which means the machine predictions can increase NPV of per-dollar investment by as much as $0.1841. Similarly, we calculate $IRR^{machine} - IRR^{crowd}$ (the gap between the "machine prediction" and "risk premium" curves in Figure 2) and find that the largest value is 0.0481, which means the maximum improvement in IRR the machine can achieve is 4.81 percentage points. The large improvement machine achieves over the crowd is partially due to the considerable negative effect that defaulted loans could have on the NPV of a portfolio.[21] Another possible reason why the crowd performs poorly in this comparison is that in reality, investors may be risk-seeking, and thus willing to invest in risky loans. We account for investors' risk preferences in the next comparison.

---

[17] The IRRs of the crowd prediction-based portfolios vary on a large scale at the beginning when only a few loans are included. To avoid the distortion of the plot, we omit the portfolios with less than 200 loans for the crowd prediction-based portfolios in the plot.

[18] The empty portfolio is not plotted.

[19] See Online Appendix A6 for portfolios for a sample of investment amounts (the values corresponding to the points in Figure 2.)

[20] The curve representing the return of loan portfolios selected based on ScoreX is above the curve representing the return of loan portfolios selected by the crowd in the right-most region (when selecting relatively high risk loans). The lower return from the crowd selected loan portfolio indicates that the crowd includes highly risky loans without asking for a sufficiently high interest rate to justify the high risk.

### Comparison (2)

In comparison (1), we show that machine-prediction-based portfolios can bring higher financial returns compared with the crowd-prediction-based portfolios. While a higher return is certainly preferred, in that comparison we have not considered crowd investors' risk preferences. It is possible that investors are risk-seeking and would prefer a riskier portfolio with a lower expected return. In comparison (2), we explicitly account for crowd investors' risk preferences. This comparison follows the contraction method proposed by Lakkaraju et al. (2017) and Kleinberg et al. (2018), which allows comparison without having complete information about decision makers' objective function (e.g., risk preferences in our case) in the presence of the selective labels problem.

To apply the method in our setting, we consider active investors in each month as "one crowd" and there are 19 crowds (decision makers) in our test set.[22] Different crowds have different risk tolerances/ preferences, which results in portfolios at different risk levels. We draw from the Modern Portfolio Theory (Markowitz, 1952), and consider the return (amount) and return variance (a risk measure) as the two components of crowd investors' utility function; the return of the portfolio equals the sum of the return on each of the loans in a portfolio, and the return variance of the portfolio equals the sum of the return variance on each of the loans in a portfolio (assuming loans are independent after controlling for observable features). As in comparison (1), we use NPV to measure the return of a loan. The return variance of a loan is calculated based on the return if the loan defaults,[23] the return if the loan is paid back, and the true probability that the loan defaults (the probability of a loan being paid back is then 1 minus the probability of the loan defaulting). As in Section 4.3.1, we bin all the funded loans into 100 quantile groups based on the machine predicted risk,[24] and use the percentage of defaulted loans in a bin as an estimate of the true default probability for the loans in that bin.

---

[21] For example, the NPV for a $5000 loan with an interest rate of 16.2% (the median interest rate in the funded set) at an annual discount factor of 2.8% (the average risk-free rate during our study period) is $1,080.06 if the loan is paid off as scheduled. If the borrower defaults on the loan after the first 12 monthly payments (the average number of payments of the defaulted loans in our dataset), the NPV of the loan would be -$2,916.37. This means that it takes about three successful loans to off-set the negative impact one defaulted loan has on the NPV.
[22] We exclude the listings that did not start and end in the same month, as they cannot be assigned to one particular "crowd".

The intuition behind our contraction method can be illustrated using the following example. Suppose that our goal is to compare the performance of the machine with that of a crowd ($j$) who invests in a portfolio with 300 listings and a return variance of $1.5 \times 10^6$. In order to see whether the machine can improve upon crowd $j$'s decision, we run the ML algorithm on the portfolio with the largest return variance, e.g., $2 \times 10^7$, and denote the crowd who constructs this portfolio as crowd $q$. The portfolio crowd $q$ invests in also tends to be larger than the portfolio crowd $j$ constructs because the more loans there are in the portfolio, the higher the return variance is. We make the machine remove loans with the highest machine predicted default risk from crowd $q$'s portfolio until the return variance of the new portfolio, denoted as $q^M$, has the same return variance as crowd $j$'s portfolio. We then compute the return (NPV) of portfolio $q^M$. This is feasible because all loans in portfolio $q^M$ are invested by crowd $q$ and therefore their labels are observable. Similarly, we compute the return of crowd $j$'s portfolio. The extent to which the return is higher for the "crowd $q$ plus the machine" case (portfolio $q^M$) is a concrete type of performance guarantee provided by the contraction process -- if the return of portfolio $q^M$ is significantly higher than that of portfolio $j$, given the two portfolios have the same return variance, we can conclude that the machine can improve upon crowd $j$'s decision. If the machine can improve upon the decisions of all crowds other than crowd $q$, we can confidently say that the machine can indeed provide performance improvement over crowd investors.

---

[23] We use the repayment rate to compute the number of payments a defaulter would make; the average number of payments among the defaulters is 12, and thus we assume the first 12 payments are made on time if a loan defaults.
[24] We also binned the loans in the test set into 100 quintiles groups based on risk premium and obtained similar results (see Online Appendix A8).

A critical condition for applying the contraction method is random assignment of the listings to the decision makers (crowds), because if listings from which crowd $q$ chooses are very different from listings from which crowd $j$ chooses, then we cannot attribute the difference in the performance (risk and return) between portfolio $q^M$ and portfolio $j$ to the difference in investment decisions the crowd and the machine make. While the arrivals of listings were not random, we leverage an empirical fact that conditional on listing amount, maximum borrower rate, credit grade and credit history length, the average listing characteristics do not appear to be systematically related to crowds' risk tolerance. Following Kleinberg et al. (2018), we divide the listings into "cells" based on the four conditioning features,[25] and focus on the 287 (out of 889 total cells) cells that contain at least five crowds with different return variances. More than 82% of the funded loans (5124 out of 6200) are captured in these 285 cells.[26] The majority of the cells are filtered out either because the number of listings in a cell is too small, or because most or all of the crowds choose not to fund the listings in the cell at all (therefore not enough variation of portfolio variance).[27] However, leaving out listings that are highly unlikely to be invested in by both the machine and the crowds will not meaningfully affect the comparison results.

To test for random assignment, we first project the outcome label (default status) onto the observed characteristics. Specifically, we regress the binary default label against all the observed features using the funded loans in the evaluation set, and then calculated the fitted values of the regression model for all the listings in the evaluation set. The fitted values essentially create an index of listing characteristics "weighted in proportion to the strength of their relationship with the outcome" (Kleinberg et al. 2018). Then within each cell, we aggregate crowds into quintiles (there are multiple crowds in each quintile group) based on their risk tolerance/preference (measured by return variance of their portfolios), and assign the quintile indicators (one indicator for each quintile except quintile 5 which serves as the baseline group) of the crowds to all the listings that are exposed to each of them. Next, separately for each cell, we regress the fitted values of the first regression model against the set of indicators for the within-cell quintile values, and calculate the p-value of the F-test statistic for the joint significance of the quintile indicators. If listing characteristics are systematically correlated with crowd portfolio variance quintiles, we would expect a concentration of low p-values. However, Figure 3 that shows the histogram of p-values across the 285 cells suggests no such concentration. Thus, we cannot reject the null hypothesis that crowd risk tolerance is not correlated with listing characteristics and that random assignment is met within each cell. We perform this test (and the contraction process we show later) at the quintile level. This is because the number of loans each crowd selects in each cell is too small.

---

[25] Specifically, we divide loan amount into sextile bins, divide credit history length into quintile bins, and divide borrower maximum rate into equal-length (0.04) bins. A cell is defined by a unique combination of loan amount quintile, credit history length quintile, borrower maximum interest rate bin, and credit grade. In order to keep more data, we merge the cells with different (binned) credit history lengths (but same other features) into larger cells if the random assignment condition can still be satisfied after the merge.

[26] Evaluation set consists of the loans in the test set and a random sample of 40% listings in the unfunded set. It can be viewed as a random sample of 40% listings from the full listing set. In Kleinberg et al. (2018), the drop in the number of observations after the authors divide cases to cells is comparable with the percentage drop in our paper.

[27] The funding rate in our dataset is 8.3%, which means the vast majority of the listings were not funded.

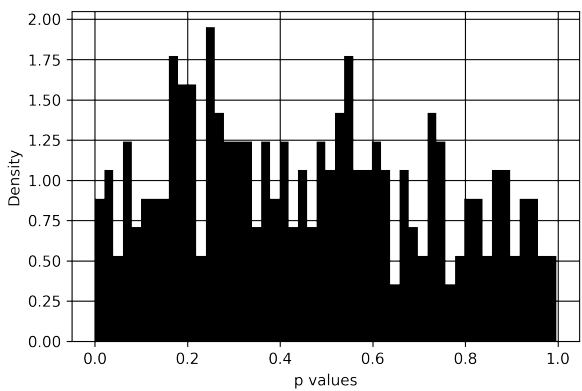

**Figure 3. Distribution of the p-values in the F-tests**

The steps of the contraction methods are as follows. We start with the portfolio with the highest return variance (the one generated by crowds in the 5th quintile) and sequentially remove loans according to their machine predicted risk from the portfolio in the highest to lowest risk sequence. As we remove an additional loan, the portfolio's return variance typically decreases, and we calculate the return for this smaller portfolio. This way, we generate a "return-variance" curve for the machine (moving from the right to the left in the left panel of Figure 4). We also plot the combination of return and variance of the five portfolios generated by the five crowd quantiles respectively. As we can see from the left panel of Figure 4, the curve (the machine's "return-variance" curve) is above the "return-variance" pair for all 1st-4th quintiles of crowds, indicating that compared with the crowd decisions, for the same portfolio variance, machine decisions always lead to higher returns (NPV in this case) except the rightmost point (i.e., the starting point of the contraction). Since no matter whether the crowd is risk-averse or risk-seeking, given the same portfolio return variance, the crowd will always prefer the portfolio that gives a higher return, this comparison suggests that machine predictions can improve upon crowd decisions after accounting for crowds' risk preferences. We also use IRR as an alternative measure of return, which is not affected by the amount of the investment, to generate the return-variance curve (the right panel of Figure 4). We can see that the machine again outperforms the crowd by generating a higher IRR at any value of portfolio variance except the rightmost point.

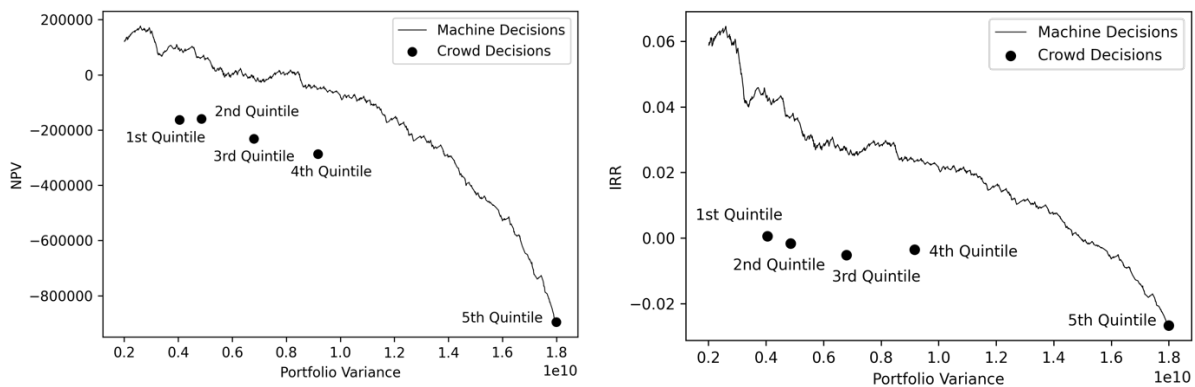

**Figure 4. Return-Variance Plot for Different Portfolios (Left Panel: NPV; Right Panel: IRR)**

**For Borrowers: More Opportunities for People in Need**

In the previous subsection, we show that the machine can increase the investor welfare. One may concern that the increase in investors' welfare is at the expense of borrowers. Borrowers care about interest rates and funding opportunities. In early years, two claimed benefits of P2P lending were the attractive interest rates and, more often emphasized, the opportunities for borrowers who are unqualified for traditional bank loans. When we use machine predictions to select loans, the interest rates remain unchanged. Therefore, machine predictions maintain this part of borrower welfare. In this section, we examine the allocation of funding opportunities under machine predictions.

Traditional banks and other funding channels usually consider several factors before making funding decisions, and credit scores are arguably the most important factor. They are often used as the first screening criterion, and people with low credit scores usually find it difficult to get funded. In addition, homeownership and credit history length are two other important factors that most lenders would consider. Borrowers with low credit scores, no homeownership and short credit histories are those who have the most difficulty getting funding through traditional channels and are most in need of alternative funding options such as P2P lending.

How are their chances under machine predictions? Similar to comparison (1)[28] in Section 4.3.2, we successively include more loans according to their machine predicted risk or risk premium ranks and then compare the borrower profiles. Figure 5 shows the change of the average ScoreX level,[29] the portions of homeowners and the average credit history as we include more loans. Except for a few extreme cases where we include less than 15% of the loans,[30] borrowers in the machine-prediction-based portfolios ($D_\phi$) have lower average credit scores, less homeownership and shorter credit histories than borrowers in the risk premium-based portfolios ($D_p$). Thus, the disadvantaged borrowers have better chances of getting funded under machine predictions.

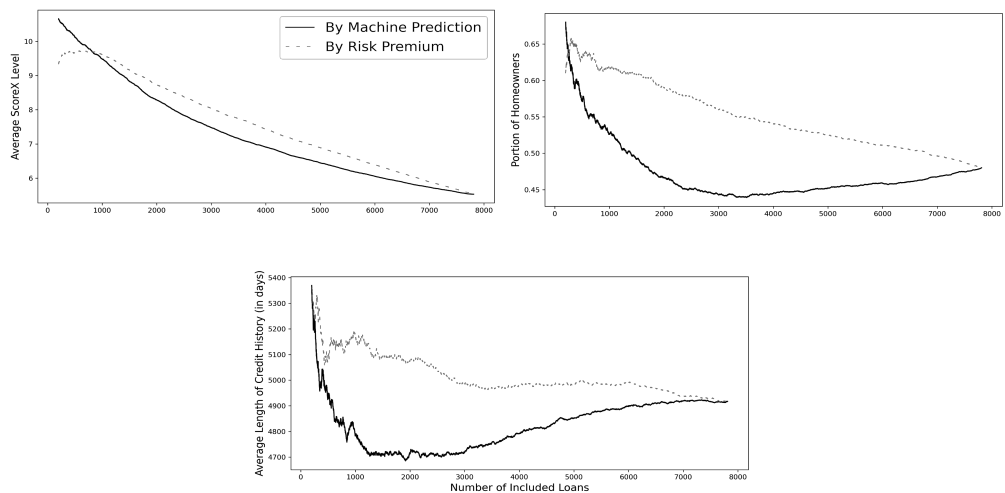

**Figure 5 Borrower Profiles as Loans Are Included**

---

[28] It is not meaningful to perform the same analysis based on loan portfolios formed using the method in comparison (2), because by construction, loans in each cell have similar values in the characteristics defining the cells (e.g., scoreX).
[29] The level mapping is as follows: {'< 600': 1, '600-619': 2, '620-639': 3, '640-649': 4, '650-664': 5, '665-689': 6, '690-701': 7, '702-723': 8, '724-747': 9, '748-777': 10, '778+': 11}.
[30] In those cases, most borrowers with relatively bad credit profiles are not considered yet, and it becomes a selection of the best borrowers among the good borrowers.

From the above analysis, we can see that machine decisions result in more opportunities to borrowers for whom P2P lending might be the most helpful. For investors, the insights about the relative *unimportance* of borrowers' credit scores, homeownership and credit history lengths in predicting loan defaults help them select investment-worthy loans and gain higher returns. Overall, we demonstrate that our ML model has greater predictive power than the crowd investors, and the increased accuracy could benefit both investors and borrowers and lead to considerable social welfare gains.

## 4.4 Why Does the Machine Outperform the Crowd?

A natural question is why the machine outperforms the crowd and for risky loans in particular. In this section, we discuss the potential reasons. One critical condition for the prediction market to work well is that participants should make their decisions independently. However, the prior literature has documented that herding behavior and social influence are common in P2P lending markets (Zhang and Liu 2012). While the observational learning (herding) may benefit individual lenders, it would harm the effectiveness of the prediction market because it breaks the independence of lenders' decisions.

However, herding behavior may explain why the machine outperforms the crowd, but it cannot explain why the improvement is larger for the relatively riskier loans. This can be explained by the design of the auction system. Recall that on Prosper, the prevailing interest rate starts from the maximum interest rate that the borrower is willing to accept. After enough investors bid on the listing to fully fund it, the prevailing interest rate starts to decrease when more investors join because they have to offer a lower rate. Note that the prevailing interest rate would never increase. Therefore, any overestimation of risk can always be corrected by later coming investors, but any underestimation of risk cannot be corrected. Thus, the high-risk loans, whose risks are more frequently underestimated, tend to be predicted with poor accuracy by the crowd.

Another reason for the underperformance of crowd predictions is that the investors may overweigh the importance of the traditional features that may indicate creditworthiness such as credit score, homeownership and length of credit history. However, the analysis reveals these are poor indicators of borrowers' creditworthiness in P2P lending context. Generally, borrowers with higher credit scores, home ownerships and long credit histories are less risky, but they also have more funding opportunities. Hence, those who choose to turn to P2P lending may be the riskier ones in their groups.

## 5. Making Fair Predictions and Decisions

As ML models start to play more important roles in a wide range of decision-making processes, people are becoming increasingly more concerned about potential algorithmic bias. Even if a ML model does not use any sensitive attributes (e.g., gender, race, and age) as inputs and the dataset contains minimal human bias, the prediction result can still be biased towards or against people with certain sensitive attribute values just because the input features are correlated with the sensitive attributes. In this section, we first present suggestive evidence for bias in the lending decisions made by the machine (i.e., the XGBoost model presented in Section 4); however, the main focus is to propose a general and effective debiasing framework that addresses several drawbacks of the existing methods. We apply it to our ML model and achieve fair predictions with a minimal (only 2%) accuracy loss.

### 5.1 Suggestive Evidence for Bias in Machine Predictions

Let $Y$ be the outcome label that is equal to 1 when the loan status is 'completed' and 0 otherwise.[31] $s$ is the predicted risk score, and $\hat{Y}$ is the predicted label.[32] As a simplified rule, the machine would choose to fund loans whose $\hat{Y} = 1$. $A$ is a binary sensitive demographic attribute.[33] $A_0$ and $A_1$ are the two sets of loans where A=0 and A=1, respectively. $X$ denotes all the explanatory variables in the dataset used as inputs to the model.

We first introduce a few commonly used fairness notions: statistical parity, equal opportunity, equalized odds and balance for the positive (negative) class. *Statistical parity* requires the probability of being funded to be the same in $A_0$ and $A_1$, i.e., $P(\hat{Y} = 1| A = 0) = P(\hat{Y} = 1|A = 1)$. It is one of the first fairness notions and is widely used in the ML community. *Equal opportunity* requires the probability of being funded to be the same for borrowers who would repay their loans in $A_0$ and $A_1$ (equal true positive rate), i.e., $P(\hat{Y} = 1|Y = 1, A = 0) = P(\hat{Y} = 1|Y = 1, A = 1)$. If in addition to *equal opportunity*, the probability of being funded is also the same for borrowers who would default in $A_0$ and $A_1$ (equal false positive rate), i.e., $P(\hat{Y} = 1|Y = 0, A = 0) = P(\hat{Y} = 1|Y = 0, A = 1)$, then the model achieves *equalized odds* (Hardt et al. 2016). Built on similar ideas, the *balance for the positive (negative) class* requires the same expected prediction score for borrowers who would payback (default on) their loans in $A_0$ and $A_1$, i.e., $E(s|Y = 1, A = 0) = E(s|Y = 1, A = 1)$ ($E(s|Y = 0, A = 0) = E(s|Y = 0, A = 1)$). These notions are closely related to the impact parity that is enforced by laws, and a prediction that violates any of them can easily be accused of being discriminatory.

---

[31] To be consistent with the fair machine learning literature and for the ease of interpretation, we choose to predict loan paybacks that would lead to a favorable outcome (funding). This is just a label flip: payback = 1-default, and it does not affect the prediction.
[32] We select a threshold for $s$ to set $\hat{Y}$ such that the number of $\hat{Y} = 1$ equals to the expected number of payback loans.
[33] All the fairness notions and debiasing methods that are described below can be easily extended to cases where the sensitive attribute has multiple classes.

Since we do not have any sensitive borrower demographic features, we use occupation and location as proxies to create gender groups and race groups (see more details in Section 3). Note that our sample size is reduced (4794 observations in the gender groups and 7716 observations in the race groups) because we cannot infer the gender or race of the borrowers for some of the loans. Additionally, since our gender proxy and race proxy are basically collections of occupations and locations, any biases we may find are also occupational biases and geographic biases. We focus on occupations and locations that are highly correlated with gender and race. Therefore, any bias against occupations/locations with a high concentration of individuals in a gender/race group provide suggestive evidence for gender/race bias. Furthermore, although using occupation and location that are highly correlated with gender and race to discriminate borrowers may not necessarily be illegal, it is still concerning. Systematic denial of financial services and resources to residents of specific, most notably black, neighborhoods or communities is called "redlining". The Community Reinvestment Act requires the Federal Reserve and other federal banking regulators to encourage financial institutions to help meet the credit needs of the communities in low and moderate-income neighborhoods, and makes open redlining in credit practices illegal. Economists have also called attention to the occupation based social segregation (Arrow 1998) as an obstacle to social inclusion.

For loans with approximated gender information, we randomly split them into a training set with 60% of the observations and a test set with 40% of the observations, train an XGBoost model using the training set, and evaluate the model performance and potential bias using the test set. We do the same for loans with approximated race information. Panel A of Table 4 shows the results for the two XGBoost models. Because the sample size shrinks substantially when we work with the subsets with approximated demographic information, the prediction accuracy suffers a bit. We still observe significant gains in returns using our models, but the disparity is also apparent. Notice that our model's inputs do not contain any gender (occupation) or race (location) information, and our prediction objective (loan default) contains minimal human bias,[34] thus, the model has no intention to discriminate on gender (occupation) or race (location) at all. Nonetheless, the prediction outcome clearly favors borrowers in a male-concentrated occupation and those living in a black-concentrated location.[35] It is difficult to accuse anyone for bias in crowd predictions, while Prosper might get into trouble when its predictive algorithm that is used for decision making is deemed to be discriminatory. This might be one of the reasons that many P2P lending platforms (e.g., Monestro, Sofi, Thincats and Flender) are still using auction-based mechanisms. In the next section, we present a general debias method, which can be used to remove/reduce bias in any feature that is deemed sensitive in any application contexts.

**Table 4. Fairness Checks**

| Panel A: Fairness Check for the Original XGBoost Models | | | | | | |
|---|---|---|---|---|---|---|
| | Gender Groups | | | Race Groups | | |
| **Number of Observations in Test Set** | **1918** | | | **3087** | | |
| AUC | 0.7173 | | | 0.7150 | | |
| | Female | Male | Mean Difference | Black | Non-Black | Mean Difference |
| Prob. Of Being Funded | 0.6013 | 0.7513 | -0.1500*** | 0.6955 | 0.5972 | 0.0983*** |
| True Positive Rate | 0.7119 | 0.8111 | -0.0992*** | 0.7857 | 0.6960 | 0.0897*** |
| False Positive Rate | 0.4201 | 0.5884 | -0.1683*** | 0.5112 | 0.4154 | 0.0957*** |
| Average Score of Positive Class | 0.7010 | 0.7713 | -0.0703*** | 0.7370 | 0.7022 | 0.0349*** |
| Average Score of Negative Class | 0.5765 | 0.6318 | -0.0553*** | 0.5966 | 0.5351 | 0.0615*** |
| Panel B: Fairness Check for the De-Biased Models | | | | | | |
| | Gender Groups | | | Race Groups | | |
| **Number of Observations in Test Set** | **1918** | | | **3087** | | |
| AUC | 0.7004 | | | 0.7123 | | |
| | Female | Male | Mean Difference | Black | Non-Black | Mean Difference |
| Prob. Of Being Funded | 0.6908 | 0.6926 | -0.0018 | 0.6713 | 0.6697 | 0.0016 |
| True Positive Rate | 0.7754 | 0.7651 | 0.0104 | 0.7619 | 0.7700 | -0.0081 |

---

[34] In some prediction problems, such as college admission predictions or judge release decision predictions, the prediction objective itself is created by humans and may contain human bias. In these cases, machines would inherit or even amplify the human bias.
[35] We get qualitatively similar results when using 15% as the percentage cutoff value when defining female/male groups and black and non-black groups (see Online Appendix A10).

|  |  |  |  |  |  |  |
|---|---|---|---|---|---|---|
| False Positive Rate | 0.5521 | 0.4952 | 0.0569 | 0.4862 | 0.4853 | 0.0009 |
| Average Score of Positive Class | 0.7392 | 0.7408 | -0.0015 | 0.7312 | 0.7382 | -0.0070 |
| Average Score of Negative Class | 0.6242 | 0.5991 | 0.0251 | 0.5926 | 0.5816 | 0.0110 |

Notes: (1) We do not directly observe gender and race in the data. We assign loans that were requested by borrowers who were in high/low female concentrated occupations (female percentage greater than 75%/less than 25%) into the female/male group; we assign loans that were requested by borrowers who were in high/low black concentrated locations (black percentage greater than 75%/less than 25%) into the black/non-black group.
(2).p<0.1; *p<0.05; **p<0.01; ***p<00.01

## 5.2 Debias Machine Prediction

The bias problem is ubiquitous in ML algorithms (Chouldechova 2018). Even though most of the algorithms do not use sensitive attributes and have no intention to discriminate, the prediction outcome would be biased when input features are correlated with sensitive attributes, which are commonly referred as *redundant encodings* (Pedreshi et al. 2008). Several methods have been proposed to help correct the bias problem (Berk et al. 2017, Calders et al. 2009, Hardt et al. 2016, Kamiran et al. 2010, Pedreshi et al. 2008). While these methods achieve promising results, and each has its own merits, they suffer from one or more of the following drawbacks: 1) can only reduce but not remove bias; 2) limited to certain ML algorithms (e.g., tree-based model); 3) limited to a certain prediction task (e.g., binary classification); 4) substantially sacrifice predictive accuracy; 5) involves changing outcome labels, which is justified in some applications where the label itself is subjective and potentially biased, but not in our case where the label is objective; and/or 6) involves setting different thresholds for different sensitive groups, which might constitute direct discrimination and its legitimacy is questionable.

Building on previous literature (Feldman et al. 2015), we propose a practical method that effectively removes bias in ML models. It is applicable to any algorithm or any type of prediction task, and in our context, results in minimal accuracy loss. The idea is to remove *redundant encodi*ng and make the input features independent of the sensitive attributes, which make it impossible for a model to infer the sensitive attributes. Formally, for each feature $X^j$ in the feature vector $X$, we find a new feature $\tilde{X}^j$ such that the following holds:

$$X^j = \sigma(\tilde{X}^j, A) \tag{10}$$

$$\tilde{X}^j \perp A \tag{11}$$

We combine $\tilde{X}^j$ to form the new feature vector $\tilde{X}$ and use it as the input for the prediction. This results in the predicted score $s = \phi(\tilde{X}) \perp A$ and, therefore, $\hat{Y} \perp A$. Below is the proof.

Since $A$ does not appear in the $\phi$ function, and $s$ is a deterministic function of $\tilde{X}$, we have $s \perp A | \tilde{X}$. Let $p$ denotes $s$'s probability density function, we have $p(s|A) = \int p(s|\tilde{x}) p(\tilde{x}|A) d\tilde{x}$. By definition, $\tilde{X} \perp A$, $p(\tilde{x}|A) = p(\tilde{x})$. Therefore, $p(s|A) = \int p(s|\tilde{x}) p(\tilde{x}) d\tilde{x} = p(s)$, which means $s \perp A$. Since $\hat{Y} = 1$ iff $s \geq s_t$, $P(\hat{Y} = 1|A) = P(s \geq s_t|A) = P(s \geq s_t) = P(\hat{Y} = 1)$, which means $\hat{Y} \perp A$.

In other words, *statistical parity* is guaranteed when the input features are independent of the sensitive attribute. To achieve *equal opportunity*, *equalized odds*, and *balance for the positive (negative) class*, we need the condition $\tilde{X} \perp A|Y$. Our method does not ensure it, but we argue that in most cases, when $\tilde{X} \perp A$, $\tilde{X} \perp A|Y$ follows. If $\tilde{X} \perp A|Y$ does not hold, it suggests that $Y$ differently "selects" in $A_0$ and $A_1$, i.e., $A$ moderates the correlation between $\tilde{X}$ and $Y$. In that case, violation of these fairness notions might be unavoidable (unless we set different thresholds or train separate models) since the underlying mapping from $\tilde{X}$ to $Y$ differs in $A_0$ and $A_1$, but our method could still effectively reduce the disparity in these measures. Additionally, when this is the case, *statistical parity* and other fairness notions cannot be achieved simultaneously, and one must be traded off for the other.

The challenge of implementing this idea arises when we have different types of variables, as is usually the case when working with real life data. Some variables are continuous, such as income; some are categorical, such as employment status; some are ordinal, such as credit grades; and some are censored, such as delinquent amounts, which has a point mass at 0 and can take continuous positive values. To construct the feature $\tilde{x}^j$ from $x^j$ and $A$, different techniques are required for different types of variables. We propose methods for each of the four types. The specific constructions and the proofs can be found in the Online Appendix A9.

### 5.3 Debias Results

Panel B of Table 4 shows the results of applying our debiasing method to the original XGBoost model. The method substantially reduces the disparity, and none of the mean differences between the individuals with a males- and females-concentrated occupation (or individuals living in a black- and non-black concentrated neighborhood) are statistically significant, which suggest that the bias, as measured by the five commonly used fairness metrics, has been effectively removed.

Fairness comes at a cost. After all, the machine now has to make predictions with the additional fairness constraint. Thus, the prediction accuracy is expected to decrease. To measure the cost of fairness, we created 500 debiased models[36] each for the gender subset and the race subset and calculate their AUC-ROCs. The average AUC-ROC for the gender subset is 0.7021 (standard deviation of 0.0038), which means that the accuracy loss is 2.12% (standard deviation of 0.53%). Similarly, for the race subset, the average AUC is 0.7110 (standard deviation of 0.0029), which translates into an accuracy loss of merely 0.5% (with a standard deviation of 0.4%). We also compare the gains in returns[37] under the debiased models and the original models; we find that the average gain in returns is 4.24% higher in the debiased models for the gender subset (standard deviation of 8.27%), and it is 0.78% higher in the debiased models for the race subset (standard deviation of 3.82%), which suggest no significant difference in the gains in returns between the debiasing models and original models. Overall, the results show that the cost of fairness is minimal in our case.

To understand why the cost of fairness in this application is minimal, we first need to identify the cause of the cost. The basic idea of our debiasing method is to make the input features independent of the sensitive attributes. In doing so, we discard information from and introduce noise to the original data, which leads to accuracy loss. There are two possible explanations for the minimal accuracy loss in our context. First, the correlations between the sensitive attributes and the other features (X variables) are weak. In this case, the amounts of discarded information and the introduced noise are small. Second, the correlations between the sensitive attributes and the outcome variable are weak. In this case, the discarded information and the introduced noise are irrelevant for predicting defaulting behavior. In our application, either or both types of correlation are weak, and thus the machine predictions are not greatly affected by the discarded information and the introduced noise, and the accuracy loss is minimal. The cost of fairness would be larger in applications where both correlations are strong.

Our debiasing method is general and not limited to the specific ML model or the investment context. Since the method directly removes the "redundant encoding" in the input data, it does not have any restrictions on the predictive model or the prediction task. It can be used in regression, classification, clustering, or even complicated ensemble models since we have shown that when $\tilde{X} \perp A$, $\hat{Y} = \phi(\tilde{X}) \perp A$ for any deterministic function $\phi(\cdot)$. In addition, the method can handle four types of variables, including continuous, categorical, ordinal and censored continuous, which cover most of the structured data types. Therefore, this debiasing method is easily applicable to datasets in other contexts where fairness may be a concern.

---

[36] Since our debiasing method introduces randomization, each time we apply it to X, we get a slightly different $\tilde{X}$; therefore, there is a slightly different de-biased model.

[37] The difference between the counterfactual returns of granting loans to those whose $\hat{Y} = 1$ and the actual returns.

## 6. Discussion and Conclusion

In this paper, we compare crowd decisions with ML algorithms in the P2P lending context. Using data from a major P2P lending platform, Prosper.com, we first show that a reasonably sophisticated ML model predicts the listing default risk more accurately than crowds of investors. Specifically, the accuracy improvement of the algorithm is greater for the listings with higher risks. We further demonstrate that the improved accuracy can lead to simultaneously higher investor welfare and borrower welfare since machine prediction can bring substantially higher returns when used for loan selection while also providing more opportunities to borrowers with few alternative funding options. We show that the machine can construct better performing portfolios in terms of NPV and IRR for investors with any risk preferences. Despite all these benefits, we recognize that there is unintended bias in machine predictions, even when sensitive attributes or attributes that can be used to infer sensitive attributes are not used as inputs. We propose a general and effective debiasing method that achieved fairness at the cost of minimal accuracy losses. After being debiased, the ML algorithm achieves fair and more accurate predictions, which considerably improves the crowd decisions in multiple aspects.

It is important to note that we do not claim that ML algorithms outperform prediction markets or crowd wisdom in general. Instead, we show the great potential of leveraging the strong predictive power of algorithms in decision making in this specific P2P lending application. In fact, we illustrate the efficiency of crowd predictions since they beat the industry standard of credit scores and are consistent with machine predictions for low risk listings. It is for high risk listings that the crowd predictions fall short. Furthermore, the underperformance of crowd predictions can be explained by the herding effect, a one-sided auction system design and credit score bias, which hinder the efficient implementation of prediction markets and are specific to the context. Similar results will likely hold on other P2P platforms where these hindrances exist; on platforms where the crowd wisdom can be better leveraged, either through a different design of the platform mechanism (e.g., a different pricing model than the one-sided auction) or through investor learning over a sufficiently long period of team (e.g., to reduce credit score bias or herding effect), ML algorithms may not necessarily outperform the crowd.

Our paper has important managerial implications. First, when crowds make investment decisions on P2P lending platforms, they tend to rely on indicators of creditworthiness that are commonly used in traditional financial institutes, such as high credit scores, home ownership and long credit histories. As a result, the borrowers with few alternative funding options do not have good chances in P2P lending either. When ML algorithms are used to replace crowd decisions, they both bring higher returns to the investors and give more opportunities to these borrowers with greater needs. Therefore, ML algorithms can benefit both investors and borrowers simultaneously and expand effective market sizes. Second, while several P2P lending platforms have shifted to ML-based posted-price models, many platforms have kept using auction-based pricing models that allow investors to make all the decisions. Our results suggest that crowds are not necessarily inferior to machines since the crowd and machine predictions are comparable for low risk listings. Nevertheless, the platforms should be careful when designing auction rules since the common auction system that only allows investors to bid *down* prices would prevent corrections for undervalued listings and therefore lead to inaccurate crowd predictions, especially for high risk listings. Third, although ML algorithms have high predictive powers, they can be biased even when algorithm designers have no intention to discriminate and demographic features are not included at all. Practitioners therefore should be cautious about using ML in fintech applications. It is difficult to accuse anyone when a crowd is biased, while a platform or a firm might get into trouble when its predictive algorithm that is used for decision making is deemed to be discriminatory. Our proposed general debiasing method provides one possible way to address the bias issue in these applications. Last, traditional decision models that heavily rely on credit scores (such as scoreX) may not be appropriate in the P2P lending context, and new models need to be developed. Our empirical results reveal that borrowers who self-select to P2P lending platforms may be different from the average borrowers, and thus credit scores provide little information about borrowers' default risk. ML models that keep evolving and updating would be more relevant and lead to better decision making.

In this paper, we view the crowd investors as a whole, compare their collective investment decisions with the machine decisions and evaluate their respective total returns. We focus on the size of total returns and leave the question of how to divide the returns among investors open. The allocation problem and competitive gaming behavior among investors are beyond the scope of this work, and they are possible future search directions. Moreover, while it is important to answer the question "can machines outperform humans?", the comparison itself should not be the end. As we increasingly use algorithms in decisions, we should also be aware of the hidden problems and unintended consequences, such as bias issues. Additionally, as Kleinberg et al. (2018) noted, predictive models can also help us understand human behaviors and errors. In addition to the "which" question, we also address the "how" and "why" questions. The ultimate goal is to make better decisions, and this can be achieved by making good use of machines, designing systems to guide human behaviors, or combining machines and humans to create synergies. This paper is an attempt along this line and to progress, we need a deeper understanding of prediction, causation and human behaviors.

stock market crisis events using deep and statistical machine learning techniques. Expert Systems with Applications, 112: 353-371.

Chen T, Guestrin C (2016) XGBoost : Reliable Large-scale Tree Boosting System. *Conf. Knowl. Discov. Data Min.*

Chen W, Lin M, Zhang BZ (2018) Lower Taxes, Smarter Crowd? The Impact of Tax Incentives on Equity Crowdfunding. *SSRN Electron. J.* https://ssrn.com/abstract=3206256.

Chouldechova A, Roth A (2018) The Frontiers of Fairness in Machine Learning. *ArXiv Prepr.* https://arxiv.org/abs/1810.08810.

Cox JC, Ingersoll Jr JE, Ross S a. (1985) A Theory of the Term Structure of Interest Rates. *Econometrica* 53(2):385–407.

Duarte J, Siegel S, Young L (2012) Trust and credit: The role of appearance in peer-to-peer lending. *Rev. Financ. Stud.* 25(8):2455–2483.

Fama EF (1970) Efficient Capital Markets: A Review of Theory and Empirical Work. *J. Finance* 25(2):383–417.

Fawcett T, Foster Provost (1997) Adaptive Fraud Detection. *Data Min. Knowl. Discov.* 316(1):291–316.

Feldman M, Friedler S, Moeller J, Scheidegger C, Venkatasubramanian S (2015) Certifying and removing disparate impact. *Proc. 21th ACM SIGKDD Int. Conf. Knowl. Discov. Data Min.* 259–268.

Gomber P, Kauffman RJ, Parker C, Weber BW (2018) On the Fintech Revolution: Interpreting the Forces of Innovation, Disruption, and Transformation in Financial Services. *J. Manag. Inf. Syst.* 35(1):220–265.

Gu S, Kelly BT, Xiu D (2018) Empirical Asset Pricing via Machine Learning. *SSRN Electron. J.* https://ssrn.com/abstract=3159577.

Hardt M, Price E, Srebro N (2016) Equality of Opportunity in Supervised Learning. *30th Conf. Neural Inf. Process. Syst.*

Herzenstein M, Sonenshein S, Dholakia UM (2011) Tell Me a Good Story and I May Lend You Money: The Role of Narratives in Peer-to-Peer Lending Decisions. *J. Mark. Res.* 48(SPL):S138–S149.

Iyer R, Khwaja AI, Luttmer EFP, Shue K (2016) Screening Peers Softly: Inferring the Quality of Small Borrowers. *Manage. Sci.* 62(6):1554–1577.

Jiang Y, Ho YC, Yan X, Tan Y (2018) Investor Platform Choice: Herding, Platform Attributes, and Regulations. *J. Manag. Inf. Syst.* 35(1):86–116.

Kamiran F, Calders T, Pechenizkiy M (2010) Discrimination aware decision tree learning. *Proc. IEEE Int. Conf. Data Min.*:869–874.

Kawai K, Onishi K, Uetake K (2014) Signaling in Online Credit Markets. *SSRN Electron. J.* https://ssrn.com/abstract=2188693.

Khandani AE, Kim AJ, Lo AW (2010) Consumer credit-risk models via machine-learning algorithms. *J. Bank. Financ.* 34(11):2767–2787.

Kleinberg J, Lakkaraju H, Leskovec J, Ludwig J, Mullainathan S (2018) Human Decisions and Machine Predictions. *Q. J. Econ.* 26:237–293.

Kleinberg J, Ludwig J, Mullainathan S, Obermeyer Z (2015) Prediction Policy Problems. *Am. Econ. Rev.* 105(5):491–495.

Lin M, Prabhala NR, Viswanathan S (2013) Judging Borrowers by the Company They Keep: Friendship Networks and Information Asymmetry in Online Peer-to-Peer Lending. *Manage. Sci.* 59(1):17–35.

Lin M, Sias RW, Wei Z (2019) The Survival of Noise Traders: Evidence From Peer-to-Peer Lending. *SSRN*37

# Online Appendices for
# Crowd, Lending, Machine, and Bias

## A1. Summary Statistics Tables

### Table A.1 Summary Statistics of the Full Dataset

|  | Full Sample | Funded Loans | Funded Loans with Matched Label | p-value |
|---|---|---|---|---|
| Sample size | 247443 | 20668 | 19,529 | - |
| **Loan characteristics** | | | | |
| Listing amount | 7879.14 | 6540.63 | 6623.36 | 0.15 |
| Listing term | 36 | 36 | 36 | - |
| Monthly payment | 290.97 | 234.00 | 236.86 | 0.16 |
| **Borrower Characteristics** | | | | |
| Stated monthly income | 4181.11 | 4637.69 | 4664.11 | 0.51 |
| Income verifiable | 0.89 | 0.94 | 0.94 | 0.93 |
| Debt to Income ratio | 0.4980 | 0.3224 | 0.3250 | 0.78 |
| Months employed | 65.60 | 68.95 | 68.97 | 0.98 |
| Employment status | | | | |
| Full-time | 0.79 | 0.86 | 0.86 | 0.76 |
| Self-employed | 0.09 | 0.08 | 0.08 | 0.79 |
| Part-time | 0.04 | 0.04 | 0.04 | 0.91 |
| Retired | 0.03 | 0.02 | 0.02 | 1.00 |
| Not employed | 0.02 | 0.01 | 0.01 | 0.88 |
| Employed | 0.03 | 0.00 | 0.00 | - |
| Other | 0.002 | 0.00 | 0.00 | - |
| Has prior Prosper loans | 0.05 | 0.11 | 0.11 | 0.62 |
| Is homeowner | 0.36 | 0.47 | 0.48 | 0.28 |
| Is Prosper lender | 0.11 | 0.25 | 0.25 | 0.49 |
| Number of Public Records (last 10 years) | 0.58 | 0.36 | 0.35 | 0.72 |
| Number of Public Records (last 12 months) | 0.06 | 0.03 | 0.03 | 0.67 |
| **Credit Characteristics** | | | | |
| Credit Grade | | | | |
| AA | 0.04 | 0.13 | 0.13 | 0.37 |
| A | 0.05 | 0.13 | 0.13 | 0.21 |

| | | | | | |
|---|---|---:|---:|---:|---:|
| | B | 0.07 | 0.17 | 0.17 | 0.25 |
| | C | 0.12 | 0.21 | 0.21 | 0.86 |
| | D | 0.17 | 0.18 | 0.18 | 0.31 |
| | E | 0.17 | 0.08 | 0.08 | 0.35 |
| | HR | 0.38 | 0.08 | 0.08 | 0.11 |
| ScoreX | | | | | |
| | < 600 | 0.23 | 0.17 | 0.16 | 0.06 |
| | 600-619 | 0.05 | 0.09 | 0.09 | 0.55 |
| | 620-639 | 0.05 | 0.09 | 0.09 | 0.43 |
| | 640-649 | 0.03 | 0.07 | 0.07 | 0.82 |
| | 650-664 | 0.03 | 0.08 | 0.08 | 0.95 |
| | 665-689 | 0.04 | 0.11 | 0.11 | 0.61 |
| | 690-701 | 0.02 | 0.05 | 0.05 | 0.71 |
| | 702-723 | 0.02 | 0.08 | 0.08 | 0.36 |
| | 724-747 | 0.02 | 0.08 | 0.09 | 0.30 |
| | 748-777 | 0.02 | 0.08 | 0.08 | 0.42 |
| | 778+ | 0.02 | 0.08 | 0.08 | 0.54 |
| | Missing | 0.46 | 0.00 | 0.00 | - |
| Current credit lines | | 8.28 | 9.58 | 9.62 | 0.56 |
| Open credit lines | | 7.29 | 8.22 | 8.25 | 0.62 |
| Bank utilization | | 0.6297 | 0.5507 | 0.5462 | 0.23 |
| Total open revolving accounts | | 5.96 | 6.36 | 6.38 | 0.68 |
| Installment balance | | 28766.14 | 26158.08 | 26131.91 | 0.94 |
| Real estate balance | | 97792.11 | 124228.62 | 126214.06 | 0.40 |
| Revolving balance | | 17890.87 | 19061.02 | 19214.45 | 0.71 |
| Total inquiries | | 11.51 | 9.04 | 9.04 | 1.00 |
| Inquiries in last 6 months | | 3.95 | 2.60 | 2.59 | 0.92 |
| Total trade items | | 25.88 | 24.60 | 24.64 | 0.82 |
| Satisfactory accounts | | 19.18 | 20.26 | 20.35 | 0.52 |
| Now delinquent derogatory | | 2.43 | 0.96 | 0.95 | 0.48 |
| Was delinquent derogatory | | 4.27 | 3.38 | 3.34 | 0.38 |
| Delinquencies over 30 days | | 9.45 | 6.67 | 6.60 | 0.52 |
| Delinquencies over 60 days | | 4.37 | 2.75 | 2.72 | 0.49 |
| Delinquencies over 90 days | | 8.74 | 4.93 | 4.86 | 0.51 |
| Amount delinquent | | 3454.80 | 1100.30 | 1080.28 | 0.75 |
| Length of credit history | | 4809.67 | 4907.40 | 4919.86 | 0.64 |
| **Outcomes** | | | | | |
| Default | | | 0.32 | 0.32 | 0.36 |
| Principal Paid | | | 4898.07 | 4936.77 | 0.44 |
| Interest Paid | | | 1341.93 | 1352.17 | 0.49 |

## Table A.2 Summary Statistics for the Gender Groups

|  | All Listings | | | Funded Loans | | |
| --- | --- | --- | --- | --- | --- | --- |
|  | Female | Male | p-value | Female | Male | p-value |
| Sample Size | 15313 | 18187 | - | 1888 | 2909 | - |
| **Loan Characteristics** | | | | | | |
| Listing amount | 6050.87 | 7734.51 | <.01 | 5282.03 | 6981.21 | <.01 |
| Listing term | 36 | 36 | - | 36 | 36 | - |
| Monthly Payment | 233.67 | 291.28 | <.01 | 192.28 | 246.94 | <.01 |
| **Borrower Characteristics** | | | | | | |
| Stated monthly income | 3101.24 | 5174.16 | <.01 | 3306.27 | 5197.60 | <.01 |
| Income verifiable | 0.94 | 0.90 | <.01 | 0.97 | 0.97 | 0.51 |
| Debt to Income ratio | 0.39 | 0.31 | <.01 | 0.34 | 0.26 | 0.06 |
| Months Employed | 63.75 | 77.12 | <.01 | 65.14 | 78.13 | <.01 |
| Employment status | | | | | | |
| Full-time | 0.90 | 0.85 | <.01 | 0.93 | 0.93 | 0.94 |
| Self-employed | 0.02 | 0.09 | <.01 | 0.01 | 0.05 | <.01 |
| Part-time | 0.05 | 0.01 | <.01 | 0.06 | 0.02 | <.01 |
| Retired | 0.00 | 0.00 | <.01 | 0.00 | 0.00 | 0.77 |
| Not Employed | 0.00 | 0.00 | <.01 | 0.00 | - | - |
| Employed | 0.03 | 0.04 | <.01 | 0.00 | 0.00 | - |
| Other | 0.00 | 0.00 | 0.54 | 0.00 | 0.00 | - |
| Has prior prosper loans | 0.09 | 0.11 | <.01 | 0.10 | 0.12 | 0.01 |
| Is homeowner | 0.33 | 0.43 | <.01 | 0.40 | 0.54 | <.01 |
| Is Prosper Lender | 0.10 | 0.19 | <.01 | 0.15 | 0.32 | <.01 |
| No. of Public Records (last 10 years) | 0.57 | 0.51 | <.01 | 0.43 | 0.33 | <.01 |
| No. of Public Records (last 12 months) | 0.06 | 0.05 | <.01 | 0.04 | 0.03 | <.01 |
| **Credit Characteristics** | | | | | | |
| Credit Grade | | | | | | |
| AA | 0.03 | 0.06 | <.01 | 0.08 | 0.16 | <.01 |
| A | 0.04 | 0.07 | <.01 | 0.09 | 0.14 | <.01 |
| B | 0.07 | 0.10 | <.01 | 0.14 | 0.18 | <.01 |
| C | 0.13 | 0.16 | <.01 | 0.22 | 0.21 | 0.58 |
| D | 0.19 | 0.19 | 0.31 | 0.23 | 0.16 | <.01 |
| E | 0.18 | 0.15 | <.01 | 0.11 | 0.08 | <.01 |
| HR | 0.36 | 0.27 | <.01 | 0.13 | 0.07 | <.01 |
| ScoreX | | | | | | |
| < 600 | 0.54 | 0.42 | <.01 | 0.23 | 0.15 | <.01 |
| 600-619 | 0.10 | 0.09 | <.01 | 0.12 | 0.07 | <.01 |
| 620-639 | 0.09 | 0.10 | 0.13 | 0.11 | 0.09 | 0.02 |
| 640-649 | 0.05 | 0.05 | 0.54 | 0.08 | 0.06 | 0.05 |
| 650-664 | 0.05 | 0.06 | <.01 | 0.09 | 0.09 | 0.83 |
| 665-689 | 0.05 | 0.08 | <.01 | 0.10 | 0.12 | 0.11 |
| 690-701 | 0.02 | 0.03 | <.01 | 0.05 | 0.05 | 0.31 |
| 702-723 | 0.03 | 0.05 | <.01 | 0.07 | 0.08 | 0.01 |
| 724-747 | 0.03 | 0.04 | <.01 | 0.06 | 0.09 | <.01 |
| 748-777 | 0.02 | 0.04 | <.01 | 0.06 | 0.09 | <.01 |
| 778+ | 0.02 | 0.04 | <.01 | 0.04 | 0.10 | <.01 |
| Current credit lines | 8.51 | 8.77 | <.01 | 9.28 | 9.74 | <.01 |
| Open credit lines | 7.47 | 7.71 | <.01 | 7.82 | 8.34 | <.01 |
| Bank utilization | 0.65 | 0.60 | <.01 | 0.58 | 0.53 | <.01 |
| Total open revolving accounts | 5.73 | 5.53 | <.01 | 6.20 | 6.11 | 0.50 |
| Installment balance | 25551.62 | 28169.07 | <.01 | 21708.11 | 24723.79 | <.01 |
| Real estate balance | 62000.01 | 99735.11 | <.01 | 80476.52 | 133178.85 | <.01 |
| Revolving balance | 11770.23 | 16468.26 | <.01 | 12104.01 | 17943.58 | <.01 |
| Total inquiries | 11.70 | 11.86 | 0.18 | 9.15 | 9.13 | 0.96 |
| Inquiries in last 6 months | 3.48 | 3.50 | 0.75 | 2.70 | 2.60 | 0.34 |
| Total trade items | 26.48 | 24.92 | <.01 | 24.53 | 24.54 | 0.97 |
| Satisfactory accounts | 18.29 | 18.66 | 0.01 | 19.08 | 20.59 | <.01 |
| New delinquent derogatory | 3.18 | 2.28 | <.01 | 1.34 | 0.78 | <.01 |
| Was delinquent derogatory | 5.00 | 3.99 | <.01 | 4.11 | 3.18 | <.01 |
| Delinquencies over 30 days | 11.37 | 8.76 | <.01 | 8.46 | 6.01 | <.01 |
| Delinquencies over 60 days | 5.55 | 3.95 | <.01 | 3.60 | 2.27 | <.01 |
| Delinquencies over 90 days | 11.10 | 7.51 | <.01 | 6.09 | 3.98 | <.01 |
| Amount delinquent | 2861.22 | 2467.65 | <.01 | 1194.10 | 921.99 | 0.07 |
| Length of credit history | 4920.65 | 4598.94 | <.01 | 4962.55 | 4734.01 | <.01 |
| **Outcomes** | | | | | | |
| Default | | | | 0.36 | 0.27 | <.01 |
| Principal Paid | | | | 3879.56 | 5499.38 | <.01 |
| Interest Paid | | | | 1214.09 | 1355.24 | <.01 |

Table A.3 Summary Statistics for the Race Groups

|  | All Listings | | | Funded Loans | | |
|---|---|---|---|---|---|---|
|  | Black | Non-Black | p-value | Black | Non-Black | p-value |
| Sample size | 47866 | 8327 | - | 5831 | 1885 | - |
| **Loan Characteristics** | | | | | | |
| Listing amount | 7085.72 | 7235.03 | 0.04 | 6483.98 | 6387.55 | 0.53 |
| Listing term | 36 | 36 | - | 36 | 36 | - |
| Monthly Payment | 270.11 | 277.74 | <.01 | 231.80 | 230.89 | 0.87 |
| **Borrower Characteristics** | | | | | | |
| Stated monthly income | 4470.06 | 3547.65 | <.01 | 4615.89 | 4365.13 | 0.01 |
| Income verifiable | 0.87 | 0.86 | <.01 | 0.95 | 0.94 | 0.08 |
| Debt to Income ratio | 0.34 | 0.38 | <.01 | 0.30 | 0.31 | 0.44 |
| Months Employed | 67.63 | 73.53 | <.01 | 70.47 | 74.06 | 0.09 |
| Employment status | | | | | | |
|    Full-time | 0.80 | 0.77 | <.01 | 0.87 | 0.84 | <.01 |
|    Self-employed | 0.09 | 0.11 | <.01 | 0.07 | 0.08 | 0.02 |
|    Part-time | 0.03 | 0.04 | 0.10 | 0.03 | 0.04 | 0.01 |
|    Retired | 0.03 | 0.04 | <.01 | 0.02 | 0.02 | 0.27 |
|    Not Employed | 0.02 | 0.02 | 0.26 | 0.01 | 0.01 | 0.33 |
|    Employed | 0.03 | 0.03 | 0.51 | 0.00 | 0.00 | - |
|    Other | 0.00 | 0.00 | 0.06 | 0.00 | 0.00 | - |
| Has prior prosper loans | 0.10 | 0.09 | 0.27 | 0.12 | 0.12 | 0.61 |
| Is homeowner | 0.41 | 0.44 | <.01 | 0.52 | 0.51 | 0.41 |
| Is Prosper Lender | 0.14 | 0.13 | 0.04 | 0.24 | 0.23 | 0.16 |
| Number of Public Records (last 10 years) | 0.52 | 0.63 | <.01 | 0.33 | 0.47 | <.01 |
| Number of Public Records (last 12 months) | 0.05 | 0.07 | <.01 | 0.03 | 0.04 | <.01 |
| **Credit Characteristics** | | | | | | |
| Credit Grade | | | | | | |
|    AA | 0.05 | 0.04 | 0.23 | 0.13 | 0.11 | <.01 |
|    A | 0.06 | 0.06 | 0.84 | 0.14 | 0.12 | 0.05 |
|    B | 0.09 | 0.09 | 0.05 | 0.18 | 0.18 | 0.78 |
|    C | 0.16 | 0.16 | 0.31 | 0.23 | 0.20 | 0.02 |
|    D | 0.19 | 0.18 | 0.24 | 0.17 | 0.19 | 0.10 |
|    E | 0.15 | 0.17 | <.01 | 0.08 | 0.10 | <.01 |
|    HR | 0.30 | 0.30 | 0.89 | 0.08 | 0.10 | <.01 |
| ScoreX | | | | | | |
|    < 600 | 0.45 | 0.47 | <.01 | 0.16 | 0.20 | <.01 |
|    600-619 | 0.09 | 0.09 | 0.88 | 0.08 | 0.09 | 0.17 |
|    620-639 | 0.10 | 0.09 | 0.07 | 0.09 | 0.09 | 0.40 |
|    640-649 | 0.06 | 0.05 | 0.02 | 0.07 | 0.07 | 0.72 |
| 650-664 | 0.06 | 0.07 | 0.55 | 0.09 | 0.07 | 0.02 |
| 665-689 | 0.07 | 0.07 | 0.53 | 0.12 | 0.12 | 0.99 |
| 690-701 | 0.03 | 0.02 | 0.02 | 0.05 | 0.05 | 0.91 |
| 702-723 | 0.04 | 0.04 | 0.54 | 0.08 | 0.08 | 0.41 |
| 724-747 | 0.04 | 0.04 | 0.17 | 0.09 | 0.07 | 0.01 |
| 748-777 | 0.03 | 0.03 | 0.31 | 0.08 | 0.07 | 0.10 |
| 778+ | 0.03 | 0.03 | 0.53 | 0.08 | 0.07 | 0.17 |
| Current credit lines | 8.75 | 8.35 | <.01 | 9.79 | 9.19 | <.01 |
| Open credit lines | 7.73 | 7.37 | <.01 | 8.40 | 7.88 | <.01 |
| Bank utilization | 0.62 | 0.62 | 0.58 | 0.55 | 0.57 | 0.08 |
| Total open revolving accounts | 5.77 | 5.48 | <.01 | 6.40 | 5.98 | <.01 |
| Installment balance | 29660.53 | 25661.92 | <.01 | 26912.95 | 24108.01 | <.01 |
| Real estate balance | 89033.60 | 86536.73 | 0.26 | 116340.17 | 116623.84 | 0.96 |
| Revolving balance | 16145.69 | 15798.75 | 0.41 | 17613.95 | 18805.47 | 0.27 |
| Total inquiries | 10.81 | 10.40 | <.01 | 8.36 | 9.44 | <.01 |
| Inquiries in last 6 months | 3.09 | 2.98 | 0.01 | 2.28 | 2.68 | <.01 |
| Total trade items | 26.38 | 25.98 | 0.01 | 25.27 | 25.13 | 0.71 |
| Satisfactory accounts | 19.31 | 18.55 | <.01 | 20.93 | 20.05 | 0.01 |
| New delinquent derogatory | 2.75 | 2.69 | 0.21 | 0.97 | 1.21 | <.01 |
| Was delinquent derogatory | 4.32 | 4.74 | <.01 | 3.37 | 3.86 | <.01 |
| Delinquencies over 30 days | 9.83 | 10.74 | <.01 | 6.93 | 7.93 | <.01 |
| Delinquencies over 60 days | 4.57 | 4.89 | <.01 | 2.81 | 3.26 | <.01 |
| Delinquencies over 90 days | 9.28 | 9.66 | 0.04 | 4.99 | 5.79 | 0.01 |
| Amount delinquent | 3123.89 | 3018.72 | 0.40 | 1124.25 | 1512.51 | 0.03 |
| Length of credit history | 4938.47 | 5041.86 | <.01 | 4954.46 | 5114.79 | 0.02 |
| **Outcomes** | | | | | | |
| Default | | | | 0.33 | 0.34 | 0.35 |
| Principal Paid | | | | 4790.35 | 4700.74 | 0.50 |
| Interest Paid | | | | 1338.45 | 1412.68 | 0.07 |

## A2. XGBoost Model

We use an XGBoost model (Chen and Guestrin, 2016) to fit a prediction function $\phi(X)$ that outputs the predicted probability of default $P(Y = 1|X)$. XGBoost is a scalable tree based boosting system that achieves state-of-the-art results in many ML challenges. An XGBoost model is a tree ensemble model that consists of multiple regression trees (also known as CARTs). Unlike normal decision trees (classification trees) that output class labels or use the portion of positive classes as the class probability, a regression tree performs a regression in each leaf node. In our case, the logistic regression is used and each tree outputs the probability of default. The final prediction of an XGBoost model is the sum of the predictions from each regression tree. Formally, the prediction for instance $i$ with feature vector $X_i$ is as follows:

$$\widehat{Y_i} = \phi(X_i) = \sum_{k=1}^{K} f_k(X_i), \quad f_k \in F$$

where $K$ is the number of regression trees, $f_k$ is the mapping function of the k-th regression tree, and $F$ is the space of regression trees.

To learn a model, we minimize an objective function that consists of the training loss and regularization term as follows:

$$L(\phi) = \sum_i l(\widehat{Y_i}, Y_i) + \sum_k \Omega(f_k)$$

$$l(\widehat{Y_i}, Y_i) = Y_i \ln(1 + e^{-\widehat{Y_i}}) + (1 - Y_i) \ln(1 + e^{\widehat{Y_i}})$$

$$\Omega(f) = \gamma T + \frac{1}{2}\lambda \|w\|_1$$

where $T$ is the number of leaves in a tree, and $w$ is the vector of the leaf weights. $\gamma$ and $\lambda$ are hyperparameters. Intuitively, we aim to balance the prediction accuracy and the model's simplicity since we minimize the sum of the training errors and the model's complexity. The above objective function cannot be optimized using traditional optimization techniques, and the model is additively and greedily trained by learning a tree and adding it to the model in one iteration. We do this greedy search in each iteration until we finish training all $K$ trees.

Tree based models are vulnerable to overfitting problems. In practice, we use several techniques to alleviate it. First, when fitting regression trees, we can specify a maximum depth so that a tree stops growing once it reaches the depth. Second, we can subsample instances and/or features and therefore create slightly different datasets for each tree. Third, after we learn a tree $f_t$, we usually shrink it when adding it to the model.

There are several hyperparameters in an XGBoost model: the number of trees $K$, the regularization terms $\gamma$ and $\lambda$, the learning rate $\epsilon$, the maximum tree depth, the instance subsample percentage and the feature subsample percentage. We tune these parameters using five-fold cross-validation on the training set. Because our hyperparameter space is relatively large and fitting the XGBoost model is computationally expensive, we choose the Bayesian Optimization instead of the more commonly used grid search or random search to search for the optimal hyperparameters. Bayesian Optimization balances the exploration-exploitation trade-off, which helps avoid getting stuck in the local minimum. In general, it is more efficient than the grid search or random search and enables us to find a near optimal set of hyperparameters with fewer trials.

Bayesian Optimization is a common method that is used for tuning hyperparameter in ML models. Hyperparameter searching is a maximization process in which we aim to find a set of parameters $V$ that maximize our objective $h(V)$ (e.g., the cross-validated model accuracy score). The challenge is that $h(V)$ is a black box function since we do not know its explicit expression or its derivatives. The evaluation is restricted to getting a response value with a certain input. Grid search and random search enumerate large sets of possible input values and independently evaluate the responses in each iteration while Bayesian Optimization learns the shape of the objective $h(V)$ from the responses over the iterations. It starts from a Gaussian Process prior that is characterized by a mean function $\mu$ and a covariance function $k$ as follows:

$$\hat{h}_0(V) \sim GP(\mu(v), k(v, v'))$$

In each iteration, the algorithm samples a data point $(V_i, h(V_i))$, adds it to the observed data set $D_{1:t}$, and produces an updated posterior function using Bayes theorem as follows:

$$P(D_{1:t}) \propto P(\hat{h}_{t-1})P(\hat{h}_{t-1})$$

The algorithm then samples a new input V that corresponds to a potentially high value of h(V) based on its current posterior function. This process is repeated until the preset number of inputs have been sampled. Then, the algorithm tries to find the optimal point of the final posterior function as the final optimal values.

## A3. Generating ROC Curves

The AUC-ROC is a common evaluation metric that is good for measuring predicted score performance when $Y$ is unbalanced, as in our case where there are more completed loans ($Y = 0$) than default loans ($Y = 1$). To calculate the accuracy, we need a binary predicted label, and it is commonly derived by setting threshold for the predicted score as follows:

$$\hat{Y}(t) = 1 \; iff \; s \geq t$$

where $s$ is the predicted score (in our case, $\phi(X|R^* = 1)$ or $p(X, Z \,|R^* = 1)$), and $t$ is the discrimination threshold. The ROC curve plots the true positive rate (TPR) against the false positive rate (FPR) as $t$ varies. When $t > max\,(s)$, no loan is classified as being in default, and thus both TPR and FPR are 0. When $t \leq (s)$, all loans are classified as being in default, and thus both TPR and FPR are 1. When the threshold is in between, the TPR and FPR take different values and result in an ROC curve. The AUC-ROC is simply the area under this curve. It also has an intuitive interpretation as the probability that a randomly selected positive instance is ranked higher than a randomly selected negative instance using their predicted score $s$. The ROC of a random guess (no predictive power) is a diagonal line that goes through (0, 0) and (1, 1), and it has a corresponding AUC of 0.5. The ROC of a perfect prediction would go through the perfect classification point (0, 1), and its AUC would be 1. Most predictive models are between the two extreme cases, and therefore have an AUC that ranges from 0.5 and 1. Figure 1 plots the ROC curves of the machine predictions $\phi(X|R^* = 1)$ and the crowd predictions $p(X, Z \,|R^* = 1)$ on the holdout test set of 7812 loans. We can see that the curve of the machine predictions strictly dominates the curve of the crowd predictions with a higher AUC of 0.7406 compared to 0.6783. This suggests that, overall, our ML model has more predictive power than the crowd.

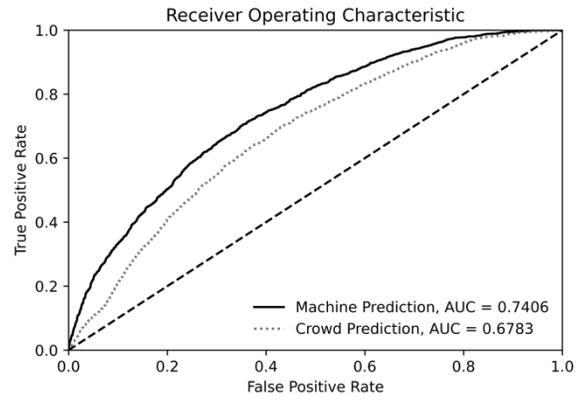

**Figure A.2 ROC Curves and AUC**

## A4. Fixed Risk-Free Rate Model

In the main analysis of comparison (1), we use risk premium as a proxy for the crowd-predicted loan default risk and assume a time-varying risk-free rate when calculating the risk premium. The rationale is that the final interest rate has two sources of variation: risk-free rate and default risk premium. Since we are interested in the crowd's assessment of the loan default risk, it is important to remove the variation in the risk-free rate from the final inters rate. However, even though the risk-free rate fluctuated daily, the variation was small, and investors may consider a fixed risk-free rate when making investment decision during a relatively short period. In this case, the risk-free rate is absorbed into the constant term, and the interest rate formula can be re-written as:

$$r(X, Z|R^* = 1) = g\big(m(X, Z|R^* = 1)\big) + c$$

This means that the only varying part of interest rate is default risk premium, and interest rate itself preserves the ordinal information in the crowd predicted default risk. Therefore, under the constant risk-free rate assumption, we can use interest rate as a proxy for the crowd prediction, and the results based on this alternative proxy are shown in Figure A.2. When assuming a fixed risk-free rate, the crowd prediction accuracy slightly increases from 0.6840 to 0.6918, but the change is minimal and our results for the main analysis (that the machine produces more accurate predictions and leads to better investment decisions) still hold.

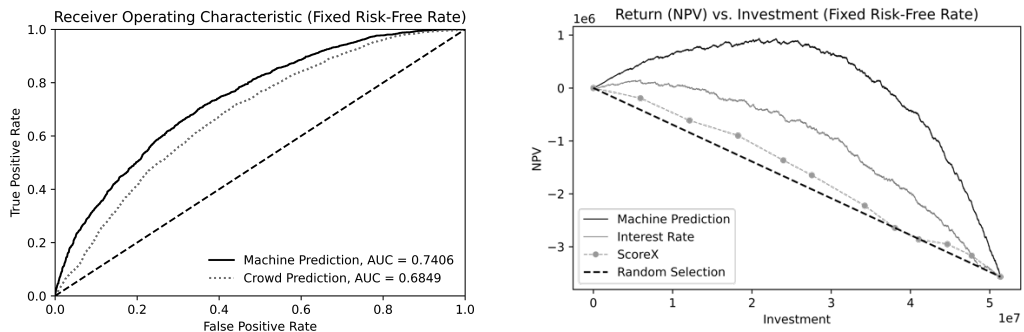

**Figure A.2 ROC Curves and Return (NPV) - Investment Plot with Fixed Risk-Free Rate**

## A5. Multiple Time Periods

In the time period of our main analysis (March 2007 to October 2008), online P2P lending platforms were considered new and becoming more and more popular, and during this period, they may have attracted borrowers and lenders who were less rational and just wanted to "give it a try". While the machine has the only goal of accurately identifying the default risks and making profitable investments, some of the crowd investors might be exploring the platform and trying things out. To examine the potential shift in lenders' and borrowers' behavior during the growth period and mitigate the concern that our results are driven by the difference in the crowd investors' objective(s) and the machine's objective, we divide our sample into 4 time periods with equal length (148 days). The period-specific summary statistics are reported in Table A.4.

**Table A.4 Summary Statistics of the Loans by Time Period**

| Period | Start Date | End Date | # of Loans | Default Rate | Avg. Interest Rate |
|---|---|---|---|---|---|
| 1 | 2007-03-01 | 2007-07-26 | 4871 | 0.3631 | 0.1720 |
| 2 | 2007-07-27 | 2007-12-22 | 4142 | 0.3421 | 0.1714 |
| 3 | 2007-12-22 | 2008-05-18 | 5336 | 0.3024 | 0.1744 |
| 4 | 2008-05-18 | 2008-10-12 | 5180 | 0.2786 | 0.1910 |

We perform the same analysis for the loans in each period as in the main analysis: split the loans into a training set and a test set, train a XGBoost model using the training data, use the model to predict default risk for loans in the test set, use risk premium as the proxy for the risk prediction of the crowd, and compare the accuracy of the predictions made by the machine and by the crowd and the decisions based on the machine and the crowd predictions. Figure A.3 and Figure A.4 show the ROC curves and the Return (NPV)-Investment plots in the four periods for comparison (1), respectively. We do not perform comparison (2) on each of the periods because the number of crowd in each period is too small.

The number of loans remained roughly the same in each period, while the default rate steadily decreased over the time. The average interest rate was stable for the first three periods and had a small jump in the last period, which might be caused by the financial crisis which started to outburst at that time. This suggests that either the crowd were getting better at identifying risky loans and becoming selective over time, or the quality of the pool of potential loans was improving. Meanwhile, the decreasing default rate means fewer positive labels (default loans), and therefore it is increasingly difficult for the machine to learn the patterns of defaulting behavior.

**Figure A.3 ROC Curves in Different Time Periods**

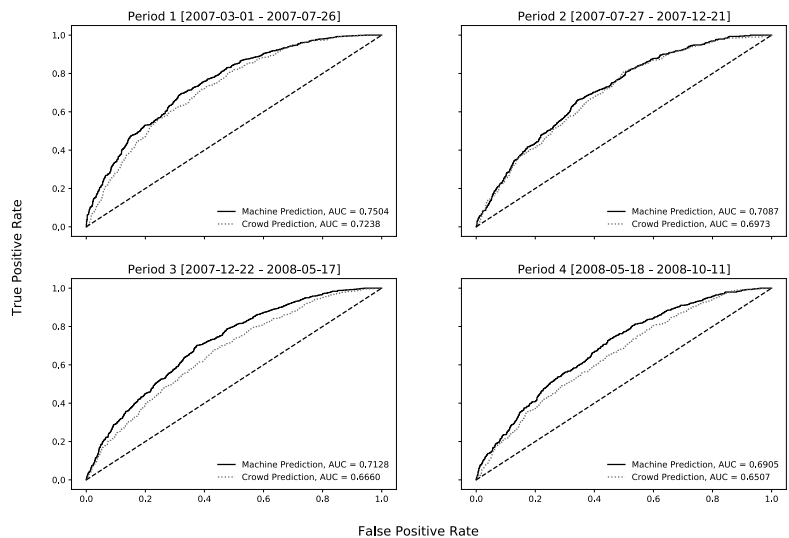

**Figure A.3 ROC Curves in Different Time Periods**

Figure A.3 and Figure A.4 show consistent results. The NPV of the loans that have been invested (in the test set) increases over time (compare the value on the vertical axis for right most point in each of the 4 plots), suggesting that the investors were improving their decisions. However, it becomes harder and harder to further identify risky loans from the selected set, and the prediction accuracy of machine prediction and crowd prediction both decline over time. It is still clear that, in all the periods, the machine makes more accurate predictions and leads to substantially higher returns. Moreover, we can see from the plots that as the crowd improved their investment decisions, the level of additional welfare gain that machine provides also increases. This may suggest that while the crowd learned to pick up certain signals to make better predictions over time, the machine may be able to capture patterns that the crowd had ignored, and therefore could generate more benefits, especially when it is difficult to identify defaulters from non-defaulters. Although Prosper might attract investors with reduced rationality or different objectives during its growth period, the overall quality of investment decisions made by the crowd noticeably improved. On top of it, the machine learning model still makes better predictions and provides an increasing level of improvement over time.

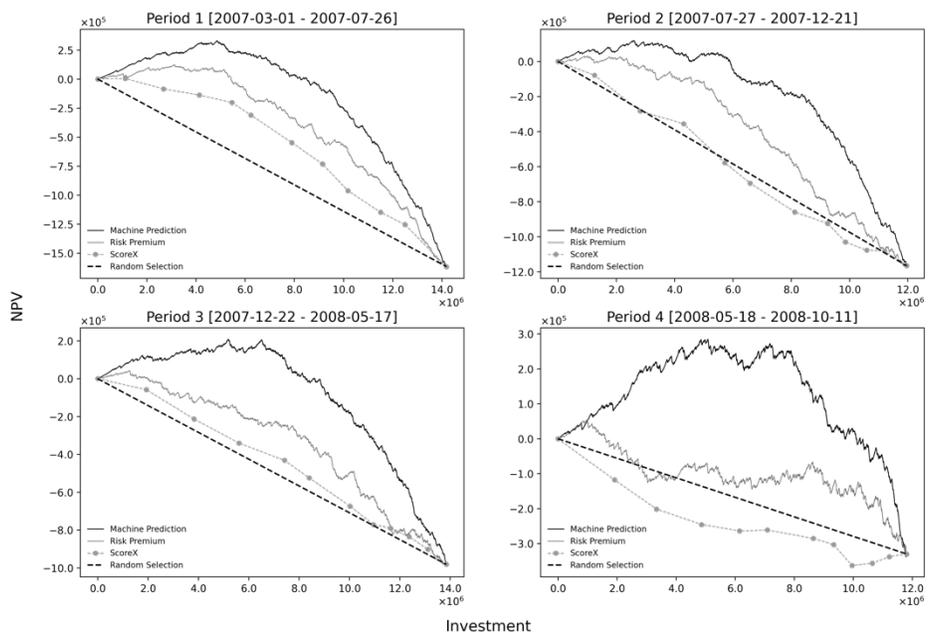

**Figure A.4 Return (NPV)-Investment Plots in Different Time Periods**

## A6. Sample Portfolios by Investment Amount

**Table A.5 Portfolios Comparison, By Investment Amount**

| Investments (in millions) | Using Machine Prediction | | Using Risk Premium | |
|---|---|---|---|---|
| | NPV | IRR | NPV | IRR |
| 1.00 | 75,614.92 | 0.0768 | 20,688.07 | 0.0416 |
| 2.50 | 207,815.02 | 0.0817 | 88,477.92 | 0.0514 |
| 5.00 | 366,299.27 | 0.0757 | 196,443.86 | 0.0540 |
| 10.00 | 653,039.22 | 0.0710 | 70,290.54 | 0.0328 |
| 15.00 | 791,477.63 | 0.0633 | -185,288.88 | 0.0195 |
| 20.00 | 881,261.76 | 0.0577 | -418,054.24 | 0.0134 |
| 25.00 | 875,970.71 | 0.0518 | -675,057.11 | 0.0090 |
| 30.00 | 625,032.66 | 0.0423 | -1,091,873.18 | 0.0022 |
| 35.00 | 170,775.40 | 0.0314 | -1,575,215.47 | -0.0042 |
| 40.00 | -423,224.07 | 0.0205 | -2,091,709.89 | -0.0096 |
| 45.00 | -1,433,494.00 | 0.0052 | -2,752,343.99 | -0.0164 |
| 50.00 | -3,050,957.47 | -0.0165 | -3,409,180.78 | -0.0219 |
| 51.34 | -3,567,424.43 | -0.0230 | -3,567,424.4 | -0.0230 |

## A7. A Rolling-Window Analysis

In the main analysis, we randomly split all the loans in the market portfolio into training and test sets and use the training set to train the machine learning model. One concern is that the machine sees the outcomes of "future loans" when making predictions for the loans in the test set, while the crowd had only access to the past information, thus the machine vs crowd comparison is not fair. To address this concern, we build a dynamic rolling-window machine learning model. Instead of randomly splitting the loans into the training and test sets, we predict default risk for each loan in the test set using the loans that started within 180 days (6 months) prior to the target loan's creation date as the training data. In this way, we ensure that the machine only has access to the past information – loans that were funded in the past 180 days – when making predictions for each loan. Such a rolling-window model also has practical value, as it is dynamically adapted when new data becomes available and therefore remains relevant for new predictions.

Since this model requires sufficient past loans as training data, we only make predictions for loans that were funded after August 2007. Figures A.5 and A.6 show the results of Comparison (1) and Comparison (2) for this rolling-window model. It suggests that our main results continue to hold under this model.

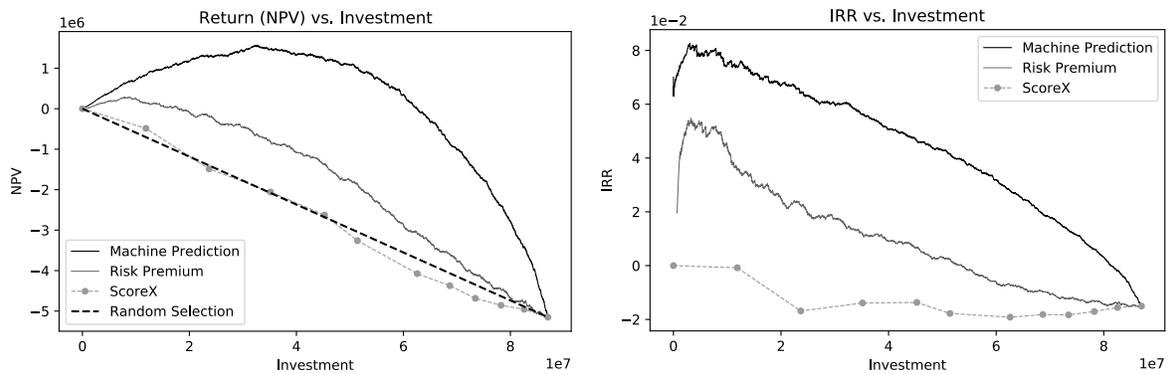

**Figure A.5 Return-Investment Plot for Different Portfolios**

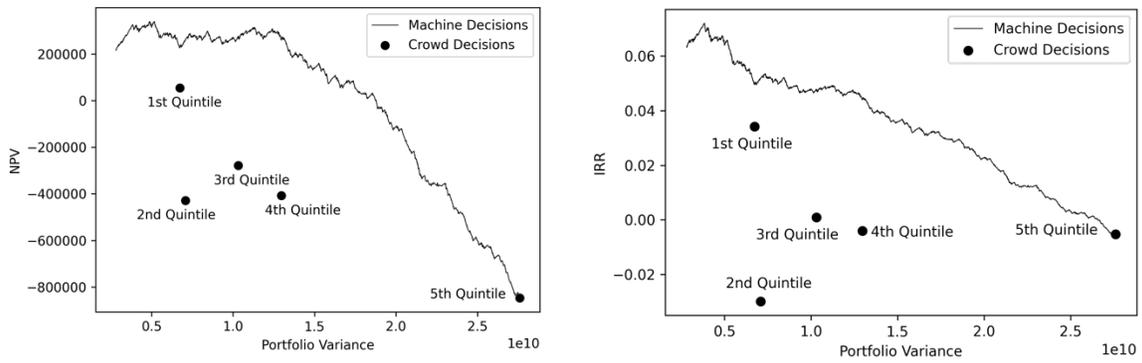

**Figure A.6 Return-Variance Plot for Different Portfolios**

## A8. Loans Binned Based on Risk Premium

In the main analysis of Comparison (2), we divide all the funded loans into 100 quantile bins based on the machine predicted risk, and use the percentage of defaulted loans in a bin as an estimate of the true default probability for the loans in that bin. In this robustness check, we define the 100 quantile bins based on the risk premium, and redo comparison (2). Figures A.7 and A.8 show the results for the random assignment check and those for the comparison. They correspond to Figures 3 and 4 in the body of the paper. As we can see from the figures, the random assignment condition still holds, and the comparison results are similar to those presented in the body of the paper.

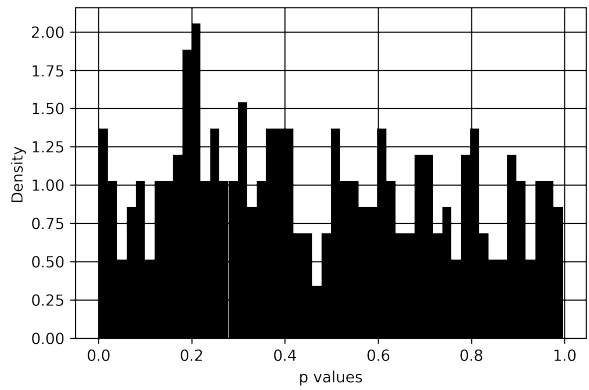

**Figure A.7 Distribution of the p-values in the F-tests**

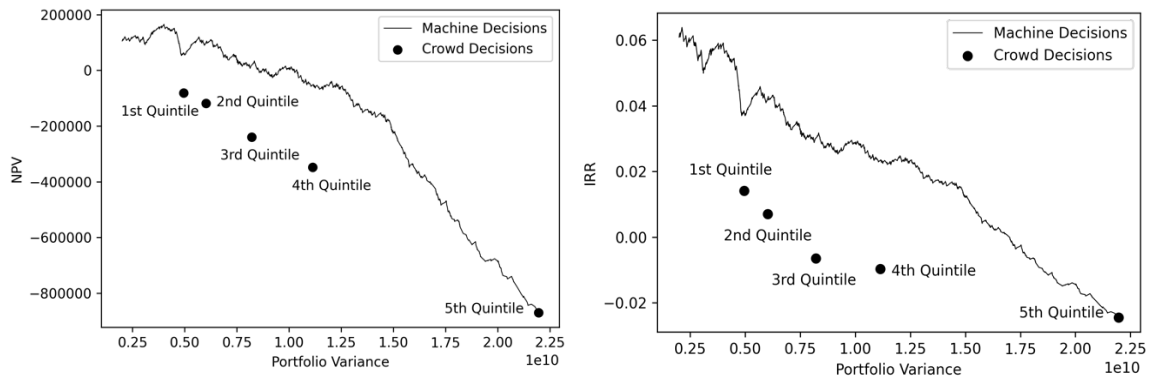

**Figure A.8 Return-Variance Plot for Different Portfolios**

## A9. Debiasing Method

**Continuous Variable**

When $X^j$ is a continuous variable, we first estimate the probability density function (PDF) of $X^j$ from the data $x_i^j$ using the kernel density estimation for each demographic group as follows:

$$\hat{f}_h(x|A = a) = \frac{1}{n_a h} \sum_{i=1}^{n} K\left(\frac{x - x_i^j}{h}\right), i \in A_a, \quad a \in \{0,1\}$$

where $n_a$ is the number of observations in each group, $K$ is the kernel function that we set to be a standard normal density function, and $h$ is a hyperparameter called the kernel bandwidth that is determined using Scott's Rule as $h = n_a^{-\frac{1}{5}}$. Let $\hat{F}_h(x^j)$ be the corresponding cumulative distribution function (CDF) as follows:

$$\hat{F}_h(x|A = a) = \int_{-\infty}^{x^j} \hat{f}_h(t|A = a)dt = P(x^j \leq x|A = a), a \in \{0,1\}$$

Then, we define a new random variable $\tilde{X}^j$ as follows:

$$\tilde{X}^j(A) = \hat{F}_h(X^j|A)$$

Each data point $\tilde{x}_i^j$, as a realization of $\tilde{X}^j$, is simply calculated as follows:

$$\tilde{x}_i^j = \hat{F}_h(x_i^j|A_i)$$

Now we prove that $\tilde{X}^j \perp A$:

$$P(\tilde{X}^j \leq x|A = 0) = P(\hat{F}_h(X^j|A = 0) \leq x)$$

$$= P(X^j \leq \hat{F}_h^{-1}(x|A = 0)|A = 0)$$

$$= x, \forall x \in [0,1]$$

Similarly:

$$P(\tilde{X}^j \leq x|A = 1) = x, \forall x \in [0,1]$$

Thus:

$$P(\tilde{x}^j \leq x|A = 0) = P(\tilde{x}^j \leq x|A = 1), \forall x \in [0,1]$$

I.e., $\tilde{x}^j \perp A$.

**Categorical (Nominal) Variable**

When $X^j$ is a categorical variable that takes $k$ values, the conditional distribution is characterized by the following:

$$P(X^j = t | A = a) = p_{at}, t = 1, \dots, k, \quad a \in \{0, 1\}$$

We define a new discrete random variable $\tilde{X}^j$ that takes $2k$ values:

$$P(\tilde{X}^j = s | A = 0) = P(\tilde{X}^j = s | A = 1) = \alpha_s, s = 1, \dots, 2k$$

The mapping function $X^j = \sigma(\tilde{X}^j, A)$ is defined as follows:

$$\begin{cases} \sigma(\tilde{X}^j = t, A = 0) = t, t = 1, \dots, k \\ \sigma(\tilde{X}^j = t + k, A = 0) = t, t = 1, \dots, k \\ \sigma(\tilde{X}^j = 2t - 1, A = 1) = t, t = 1, \dots, k \\ \sigma(\tilde{X}^j = 2t, A = 1) = t, t = 1, \dots, k \end{cases}$$

We then solve for $\alpha_s (s = 1, \dots, 2k)$ that satisfies the following conditions:

$$\begin{cases} \alpha_t + \alpha_{t+k} = p_{0t}, t = 1, \dots, k \\ \alpha_{2t-1} + \alpha_{2t} = p_{1t}, t = 1, \dots, k \\ \alpha_s \geq 0, s = 1, \dots, 2k \end{cases}$$

The system of equations is guaranteed to have solution when $k = 2$, but not for more categorical values. When the system is not solvable, we can either choose to solve the equivalent least squares problem[38] or group some of the categories to decrease the $k$.

$\tilde{X}^j$ is by design independent of A, and each data point $\tilde{x}_i^j$, as a realization of $\tilde{X}^j$, is calculated as follows:

$$\tilde{x}_i^j = \sigma^{-1}(x_i^j, A)$$

**Ordered Categorical (Ordinal) Variable**

---

[38] In that case, $\tilde{x}^j$ may be weakly dependent of A. The bias can be reduced but not removed.

When $X^j$ is an ordered categorical variable, we do not want to treat it as a nominal variable since it would lose the ordinal information during the process. Instead, we first convert the data into a continuous variable by randomly sampling value from a corresponding range as follows:

$$\hat{x}_i^j \sim unif(l_t, h_t), if\ x_i^j = t$$

where $l_t$ and $h_t$ are respectively the low and high values corresponding to the level $t$. It can be decided based on the nature of $X^j$. For example, if $X^j$ is discretized bins, $l_t$ and $h_t$ can be the natural bin ranges. If $X^j$ only takes a few sequential values, we can set the following:

$$l_t = \frac{2t-1}{2}, h_t = \frac{2t+1}{2}$$

Afterwards, we use the debiasing method for continuous variables that was described before to create $\tilde{X}^j$ from $\hat{X}^j$.

**Mixed Variable**

Some variables are continuous by nature but may have a point mass at a certain value due to missing data or censoring. For example, outstanding balance of prior prosper loans should be considered as a continuous variable, but because most the borrowers in our dataset do not have prior prosper loans, this field is 0 for them. We call such a variable a 'mixed variable' since it is a mix of continuous variables and binary variables. None of the methods described above can effectively create a new variable that is independent of $A$ from a mixed variable, and we need a special way to handle it.

Let $X^j$ be a mixed variable that has a point mass at $a$ and a continuous distribution elsewhere. We define a binary variable $u$ as follows:

$$u = 1 \text{ iff } X^j = a$$

Let $\tilde{u}$ be a debiased variable that we created from $u$, which is constructed to be a nominal variable. Therefore, $\tilde{u} \perp A$. Let $\tilde{w}$ be a debiased variable that we created from $X^j|u = 0$, which is constructed to be a continuous variable. It follows that $\tilde{w} \perp A|u = 0$.

Now, we define variable $\tilde{X}^j$ as follows:

$$\tilde{X}^j(\tilde{u}, A) = \begin{cases} \widetilde{w}(\tilde{u}, A), & if \sigma(\tilde{u}, A) = 0 \\ \widetilde{w}(\tilde{u}, 1 - A), & o.w. \end{cases}$$

By construction, $\tilde{X}^j \perp A \mid \tilde{u}$, and thus $P(A|\tilde{X}^j, \tilde{u}) = P(A|\tilde{u})$. Since $\tilde{u} \perp A$, $P(A|\tilde{u}) = P(A)$. Therefore, we have $P(A|\tilde{X}^j, \tilde{u}) = P(A)$, i.e., $A \perp \tilde{X}^j, \tilde{u}$, which implies that $\tilde{X}^j \perp A$.

## A10. Robust Check Results for Fairness Check

In the main analysis, we assign loans that were requested by borrowers who were in high/low female concentrated occupations (female percentage greater than 75%/less than 25%) into the female/male group; we assign loans that were requested by borrowers who were in high/low black concentrated locations (black percentage greater than 75%/less than 25%) into the black/non-black group. The following table shows the results when the cutoff is set to 85% and 15%.

**Table A.7 Fairness Checks**

| | Panel A: Fairness Check for the Original XGBoost Model | | | | | |
|---|---|---|---|---|---|---|
| | Gender Groups | | | Race Groups | | |
| **Number of Observations** | 1918 | | | 3087 | | |
| AUC | 0.7043 | | | 0.7348 | | |
| | Female | Male | Mean Difference | Black | Non-Black | Mean Difference |
| Prob. Of Being Funded | 0.6054 | 0.6919 | -0.0865** | 0.7019 | 0.5898 | 0.1121*** |
| True Positive Rate | 0.7021 | 0.7746 | -0.0725* | 0.8041 | 0.6747 | 0.1294*** |
| False Positive Rate | 0.4340 | 0.4936 | -0.0596 | 0.4873 | 0.4194 | 0.0680 |
| Average Score of Positive Class | 0.6934 | 0.7245 | -0.0311** | 0.7508 | 0.7078 | 0.0430*** |
| Average Score of Negative Class | 0.5458 | 0.5964 | -0.0506* | 0.5959 | 0.5514 | 0.0445* |
| | Panel B: Fairness Check for the De-Biased Model | | | | | |
| | Gender Groups | | | Race Groups | | |
| **Number of Observations** | 1918 | | | 3087 | | |
| AUC | 0.6927 | | | 0.7310 | | |
| | Female | Male | Mean Difference | Black | Non-Black | Mean Difference |
| Prob. Of Being Funded | 0.6553 | 0.6641 | -0.0088 | 0.6748 | 0.7051 | -0.0303 |
| True Positive Rate | 0.7589 | 0.7406 | 0.0183 | 0.7781 | 0.7992 | -0.0211 |
| False Positive Rate | 0.4717 | 0.4807 | -0.0090 | 0.4581 | 0.5161 | -0.0580 |

| Average Score of Positive Class | 0.7134 | 0.7133 | 0.0001 | 0.7384 | 0.7465 | -0.0081 |
| Average Score of Negative Class | 0.5720 | 0.5922 | -0.0203 | 0.5818 | 0.5974 | -0.0156 |

Notes: (1) We do not directly observe gender and race in the data. We assign loans that were requested by borrowers who were in high/low female concentrated occupations (female percentage greater than 85%/less than 15%) into the female/male group; we assign loans that were requested by borrowers who were in high/low black concentrated locations (black percentage greater than 85%/less than 15%) into the black/non-black group.

(2) .p<0.1; *p<0.05; **p<0.01; ***p<00.01